\documentclass[twoside,twocolumn,9pt]{article}
\usepackage{extsizes}
\usepackage[super,sort&compress,comma]{natbib} 
\usepackage[version=3]{mhchem}
\usepackage[left=1.5cm, right=1.5cm, top=1.785cm, bottom=2.0cm]{geometry}
\usepackage{balance}
\usepackage{mathptmx}
\usepackage{sectsty}
\usepackage{graphicx} 
\usepackage{lastpage}
\usepackage[format=plain,justification=justified,singlelinecheck=false,font={stretch=1.125,small,sf},labelfont=bf,labelsep=space]{caption}
\usepackage{float}
\usepackage{fancyhdr}
\usepackage{fnpos}
\usepackage[english]{babel}
\addto{\captionsenglish}{%
  
}
\usepackage{array}
\usepackage{droidsans}
\usepackage{charter}
\usepackage[T1]{fontenc}
\usepackage[usenames,dvipsnames]{xcolor}
\usepackage{setspace}
\usepackage[compact]{titlesec}
\usepackage{hyperref}

\definecolor{cream}{RGB}{222,217,201}

\begin{document}

\pagestyle{fancy}
\thispagestyle{plain}
\fancypagestyle{plain}{
\renewcommand{\headrulewidth}{0pt}
}

\makeFNbottom
\makeatletter
\renewcommand\LARGE{\@setfontsize\LARGE{15pt}{17}}
\renewcommand\Large{\@setfontsize\Large{12pt}{14}}
\renewcommand\large{\@setfontsize\large{10pt}{12}}
\renewcommand\footnotesize{\@setfontsize\footnotesize{7pt}{10}}
\makeatother

\renewcommand{\thefootnote}{\fnsymbol{footnote}}
\renewcommand\footnoterule{\vspace*{1pt}%
\color{cream}\hrule width 3.5in height 0.4pt \color{black}\vspace*{5pt}} 
\setcounter{secnumdepth}{5}

\makeatletter 
\renewcommand\@biblabel[1]{#1}            
\renewcommand\@makefntext[1]%
{\noindent\makebox[0pt][r]{\@thefnmark\,}#1}
\makeatother 
\renewcommand{\figurename}{\small{Fig.}~}
\sectionfont{\sffamily\Large}
\subsectionfont{\normalsize}
\subsubsectionfont{\bf}
\setstretch{1.125} 
\setlength{\skip\footins}{0.8cm}
\setlength{\footnotesep}{0.25cm}
\setlength{\jot}{10pt}
\titlespacing*{\section}{0pt}{4pt}{4pt}
\titlespacing*{\subsection}{0pt}{15pt}{1pt}

\fancyfoot{}
\fancyfoot[RO]{\footnotesize{\sffamily{~\textbar  \hspace{2pt}\thepage}}}
\fancyfoot[LE]{\footnotesize{\sffamily{\thepage~\textbar\hspace{3.45cm}}}}
\fancyhead{}
\renewcommand{\headrulewidth}{0pt} 
\renewcommand{\footrulewidth}{0pt}
\setlength{\arrayrulewidth}{1pt}
\setlength{\columnsep}{6.5mm}
\setlength\bibsep{1pt}

\makeatletter 
\newlength{\figrulesep} 
\setlength{\figrulesep}{0.5\textfloatsep} 

\newcommand{\topfigrule}{\vspace*{-1pt}%
\noindent{\color{cream}\rule[-\figrulesep]{\columnwidth}{1.5pt}} }

\newcommand{\botfigrule}{\vspace*{-2pt}%
\noindent{\color{cream}\rule[\figrulesep]{\columnwidth}{1.5pt}} }

\newcommand{\dblfigrule}{\vspace*{-1pt}%
\noindent{\color{cream}\rule[-\figrulesep]{\textwidth}{1.5pt}} }

\makeatother


\twocolumn[
  \begin{@twocolumnfalse}
  \noindent\LARGE{\textbf{Sequential Design of Adsorption Simulations in Metal-Organic Frameworks}} \\
 \vspace{0.3cm} \\

 \noindent\large{Krishnendu Mukherjee, Alexander W. Dowling, and Yamil J. Col\'{o}n$^{\ast}$} \\

 \noindent\normalsize{The large number of possible structures of metal-organic frameworks (MOFs) and their limitless potential applications has motivated molecular modelers and researchers to develop methods and models to efficiently assess MOF performance. Some of the techniques include large-scale high-throughput molecular simulations and machine learning models. Despite those advances, the number of possible materials and the potential conditions that could be used still pose a formidable challenge for model development requiring large data sets. Therefore, there is a clear need for algorithms that can efficiently explore the spaces while balancing the number of simulations with prediction accuracy. Here, we present how active learning can sequentially select simulation conditions for gas adsorption, ultimately resulting in accurate adsorption predictions with an order of magnitude less number of simulations. We model adsorption of pure components methane and carbon dioxide in Cu-BTC. We employ Gaussian process regression (GPR) and use the resulting uncertainties in the predictions to guide the next sampling point for molecular simulation. We outline the procedure and demonstrate how this model can emulate adsorption isotherms at 300 K from 10\textsuperscript{-6} to 300 bar (methane)/100 bar (carbon dioxide). We also show how this procedure can be used for predicting adsorption on a temperature-pressure phase space for a temperature range of 100 to 300 K, and pressure range of 10\textsuperscript{-6} to 300 bar (methane)/100 bar (carbon dioxide).} 
 \\

 \end{@twocolumnfalse} \vspace{0.6cm}
]


\renewcommand*\rmdefault{bch}\normalfont\upshape
\rmfamily
\section*{}
\vspace{-1cm}


\footnotetext{$^{\ast}$\textit{$^{a}$~Department of Chemical and Biomolecular Engineering, University of Notre Dame, IN 46556, USA. E-mail: ycolon@nd.edu}}




\section{Introduction}

Metal-organic frameworks (MOFs) are crystalline nanoporous materials comprised of inorganic nodes connected by organic linkers.\cite{Kondo1997} The chemical versatility of the building blocks provides a unique opportunity to tailor and design these materials with desired textural and chemical properties. The design flexibility of MOFs has resulted in their deployment for energy storage, catalysis, drug delivery, photonics, sensors, etc.\cite{Langmi2014,CO2,HU_CO2,mof_catalysis,separation,mof_sensors,mof_drug,mof_drugs,photon} Despite the potential of these materials and their increasing numbers in experimental and synthetic studies, there is a challenge to determine which are the best materials and what are the conditions (e.g., temperature, pressure) that maximize their performance.

Molecular simulations have played an important role in the design and discovery of MOFs in a variety of applications.\cite{arni} Molecular models that describe the interactions between the materials and adsorbents of interest have been used to provide important physical insights and guide experiments towards promising candidates. The number of MOFs has kept increasing and so new algorithms and techniques have been introduced to enhance computational screening capabilities.\cite{moghadam,toolbox,ms_mofs,ms_Co2_methane} Some of these algorithms include those for crystal generation and enumeration, characterization of porous structures, and performance evaluation.\cite{core,characterization,diego_bottom_up} The use of these large-scale, high-throughput computational screening techniques on databases of MOF structures (experimental or computationally generated) has revealed structure-property relationships and identified top performing materials for many applications.\cite{wilmer,mof_hts,song_hts,wollman} These studies can produce large amounts of data relating the physical and textural properties of MOFs (void fraction, surface area, pore volume, etc.) to their performance. 

The deluge of data has allowed researchers to employ machine learning (ML) algorithms in a multitude of settings, with emphasis on gas adsorption and separations.\cite{Krish,ml_shi,co2_fer,agaji,Chung_ml_co2} Some examples include hydrogen storage, methane storage, and Xe/Kr separations.\cite{h2_ml,bobbitt,methane_ml,ML_methane_fanour,xekr} These ML models have resulted in important physical insight through the development of new descriptors capable of capturing important factors for applications of interest.\cite{geometrical,rapid_ML_co2,energy-based-descriptor,eigen_cages}  ML studies have also resulted in surrogate models capable of calculations that are orders of magnitude faster than the molecular simulations they rely upon for data.\cite{befort2021machine} Therein lies a challenge and bottleneck for workflows that rely on ML for predictions: large datasets are needed for the proper training and use of many ML algorithms. In cases where obtaining data is difficult or time consuming, the potential of ML algorithms and workflows is severely limited.
As such, recent efforts have focused on using data that is already available in the literature or can be easily obtained using simulations to make predictions for new systems. We recently demonstrated this type of approach using transfer learning.\cite{TL_colon} Transfer learning leverages information used to train a model to produce a new model applied in a novel context using significantly less data. We trained deep neural networks (DNNs) for hydrogen adsorption at 243 K and 100 bar. We then used it as a source task where all the layers of the DNN remain fixed and only the last layer is fit for a new target task. New target tasks included hydrogen and methane adsorption at different temperatures. Interestingly, although the transfer learning model used an order of magnitude less data, we found higher accuracy compared with direct training. However, transferring the learning from pure component adsorption of hydrogen or methane to separations of Xe/Kr proved challenging because the underlying features that account for the behaviors are different. 

Another approach involves training a multilayer perceptron (MLP) on alchemical species; these are modeled using arbitrary forcefield parameters that do not necessarily correspond to real molecules. With enough sampling in the alchemical space, the parameters that correspond to the real molecules will be included. Anderson and coworkers successfully demonstrated this type of approach, training an MLP using isotherms of alchemical species and making accurate isotherm predictions of real and simple molecules.\cite{anderson} Extrapolations to other molecules not included in the training set showed reasonable accuracy.  Most recently, Sturluson and coworkers implemented an algorithm to complete missing adsorption and physical property data in covalent organic frameworks (COFs) based on available data.\cite{arni_cofs} They trained a low rank model of adsorption-property matrices which makes ``recommendations'' in places where there is missing data. Through this the researchers were able to make predictions of missing values and group materials by their adsorption performance.
	
Alternative approaches employ an active learning (AL) approach --- also known as sequential design --- to help balance the accuracy of the predictive models with the number of data points to be acquired. This can be particularly attractive in situations where the feature space is very large (adsorption while varying temperature and pressure conditions) and/or time-consuming or resource-intensive experiments or simulations are needed. These approaches are increasing in popularity in the molecular simulation space. Uteva and coworkers recently implemented AL for intermolecular potential energy surfaces, showing improvement over grid-based approaches.\cite{active_learning} Similarly, Vandermause used AL to balance the use of quantum mechanical calculations to produce force fields.\cite{vandermause} In the context of porous materials, Santos and coworkers present a recent example where they seek to connect different length and time scales.\cite{santos_al} To do so, they require expensive molecular dynamics (MD) simulations. They used AL where the simulations were chosen based on model uncertainty through a query-by-committee approach. They show they require an order of magnitude less simulations to build their data set.
	
Herein we present an AL approach to balance model prediction accuracy with the number of simulations required to build a reasonable data set. The method relies on Gaussian process regression (GPR) where a data prior is fit.\cite{gaussianprocesses} The GPR model returns a prediction mean and prediction standard derivation (uncertainty), the later of which is used to determine the next individual simulation to be performed. We demonstrate this approach modeling adsorption of pure components methane and carbon dioxide in Cu-BTC for single isotherms and the temperature-pressure space. We outline the algorithm and show an order of magnitude saving on the number of simulations required to accurately assess the adsorption landscape.

\section{Methods}
\subsection{Active Learning}
The procedure outlined in this work intelligently selects the next adsorption simulation to be performed to facilitate training an accurate Gaussian process (GP) surrogate model. A GP is a non-parametric ML model that describes a real process $f(x)$ with a distribution over functions which have a joint Gaussian distribution described by a mean $\mu(x)$ and covariance $K(x,x^{'})$ function\cite{gaussianprocesses}:

\begin{equation}
  f(x) \sim N(\mu(x),K(x,x^{'})).
\end{equation}

There are many potential choices for $K(x,x^{'})$. We chose the rational quadratic kernel as it has been used before to describe adsorption loading in MOFs\cite{gopalan}:
\begin{equation}
  K(x,x^{'}) = \left(1 + \frac{d(x,x^{'})^2}{2\alpha l^2}\right)^{-\alpha},
\end{equation}

\begin{figure*}
\centering
\includegraphics[height=8.0cm]{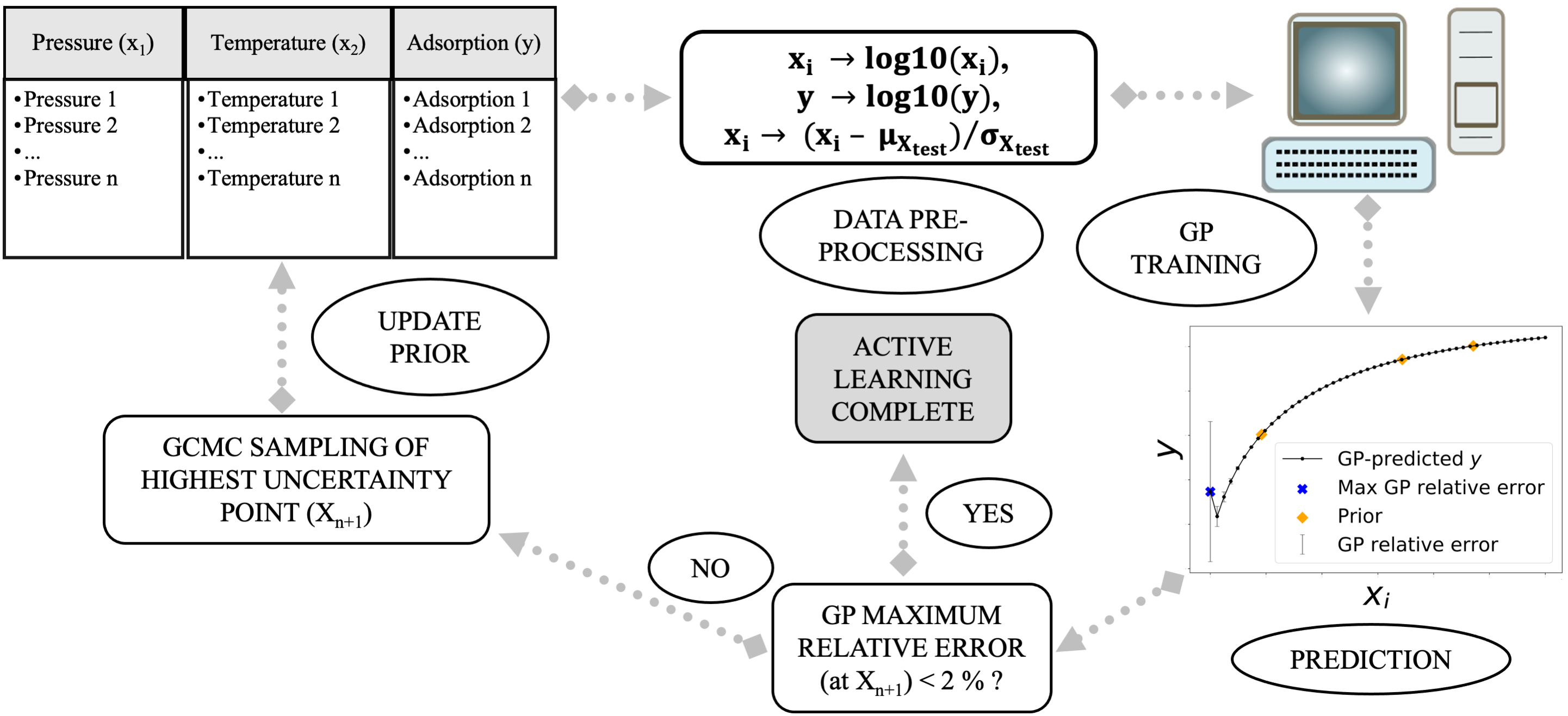}
\caption{A simple AL workflow for predicting adsorption isotherm in MOFs. The first step is generating prior data, as shown in the top left table with input variables, $x_1$, $x_2$ as pressure and temperature, and adsorption (output) as $y$. In the next step, data pre-processing is done by taking the log (base 10) for all the $x_{i}$ and $y$, followed by standardizing the input variables $x_i$. Also, the input variables are standardized with respect to the mean and standard deviation of the test set $X_{test}$. This is followed by training the pre-processed data with a Gaussian process (GP) regression. After training is complete, adsorption predictions are made for the test set. The GP predicted relative error is then calculated for all the test points on the isotherm, and then maximum GP relative error is extracted. If the value of this error is less than 2 \% (our convergence limit for the AL, except section 3.4 where it is set to 3 \%) then learning is complete. If not, then the point with maximum relative error is sampled using another GCMC simulation, and the prior is updated with this data. After prior updating, next cycle of AL begins and it goes on until the maximum GP relative error goes below threshold. 
}
\label{fgr:al_workflow}
\end{figure*}

\noindent where $d(x,x^{'})$ is the Euclidean distance between $x$ and $x^{'}$, $l$ is the length scale of the kernel, and $\alpha$ is the scale mixture parameter. The hyperparameters of the kernel, $l$ and $\alpha$, are found by maximizing the log-marginal-likelihood; the L-BGFS-B optimization algorithm implemented in scikit-learn was used in this work.\cite{lbgfs,scikit-learn} Importantly, to ensure the GP is fit appropriately, we take the log (base 10) of all the data (pressure, temperature, adsorption loading) and standardize the input variables (pressure and temperature), before it is run through the GP workflow.

Another important aspect of AL procedures is the acquisition function: how to choose the next simulation. The purpose of this study is to explore the pressure and temperature conditions quickly and accurately for a given adsorption process. To achieve this, we settled on a “greedy” approach or one that simply “explores” the space. So, for iteration $n+1$, we choose conditions (pressure or temperature and pressure) $x_{n+1}$ that maximizes the GPR prediction variance $\sigma_n^2$ constrained by bounds represented as the set $\mathcal{X}$:
\begin{equation}
    x_{n+1} = \underset{x \in \mathcal{X}}{\text{argmax}} \quad \sigma_n^{2}(x).
\end{equation}

This is known as active learning MacKay, which was originally proposed in the context of neural networks.\cite{al_macay} Seo and coworkers implemented the idea for GPs.\cite{seo_gp} After the new simulation at $x_{n+1}$ is performed and the data is gathered, the GP is refit and the procedure is repeated until $\sigma_n^2\left(x\right)$ is below some threshold. We picked this threshold as 2 \% for our methane and carbon dioxide adsorption (section 3.1, 3.2, and 3.3) while a 3 \% limit was chosen for carbon dioxide adsorption with two features (section 3.4). At the beginning of the procedure a GP prior is usually fit using data spread in $x$. Figure \ref{fgr:al_workflow} summarizes the AL procedure in this work. Also, irrespective of the performance of the AL at the first iteration, we forced the algorithm to complete the first cycle. This was done because for some specific choice of priors, the GP can become overconfident and the GP predicted relative error might be too low.
The test set for AL was linearly spaced between the pressure limits. For methane this range was 10\textsuperscript{-6} to 300 bar, while for carbon dioxide the limit was set to 10\textsuperscript{-6} to 100 bar. We used this as a test set, denoted by $X_{test}$, which was an array consisting of 50 grid points. The next point $x_{n+1}$ for AL was determined from this set only. For the case of two features (section 3.3 and 3.4), we added a temperature grid as well. The temperature test set was also linearly spaced from 100 to 300 K for both methane and carbon dioxide, and it consisted of 40 points. While we used this linearly spaced grid criteria for testing and building the AL model, we also did an interpolation test at the low pressure region (10\textsuperscript{-6} to 1 bar) for both methane and carbon dioxide. For this test, we had 50 grid points spaced in the natural log-scale to test the performance of the final GP regression after AL has finished. We only did this interpolation test for low pressure region and we kept the temperature range same as the $X_{test}$. Also for the interpolation test, the input variables were standardized against $X_{test}$. This was done to create an environment in which a user can test the power of a final AL fit model which is completely blind to the interpolation test information. The AL performance for both the AL initial test set (i.e. on $X_{test}$) and low pressure interpolation test are reported in results section. Also, a set of GCMC simulation were done for both tests to generate the ground truth data. The details of GCMC simulation uncertainty ($\sigma_{GCMC}$) for both the $X_{test}$ and low pressure interpolation test are given in the respective tables in the next sections.

\subsection{Molecular Simulations}
Adsorption loading at various temperatures (100 to 300 K) and pressures (10\textsuperscript{-6} to 300 bar) were calculated using grand canonical Monte Carlo (GCMC) simulations in RASPA. \cite{gcmc,gcmc2,raspa} MC moves employed were insertions, deletions, reinsertions at a random point in the space, rotations, and translations with equal probability; 2,000 initialization and 20,000 production cycles were used for methane and 2,000 initialization cycles and 20,000 production cycles for carbon dioxide. Methane and carbon dioxide were modeled using TraPPE.\cite{trappe} Nonbonded interactions for methane and carbon dioxide were modeled as a Lennard-Jones (LJ) or LJ + Coulomb, respectively with a cutoff for van der Waals interactions of 12.5 {\AA}.
MOF Cu-BTC with charges was chosen for this study.\cite{Lennard_Jones_1931} It was modeled as rigid and parameters for nonbonded interactions were taken from the Universal Force Field (UFF).\cite{UFF} 
Lorentz-Berthelot mixing rules were used for cross-term interactions.\cite{lorentz}

\subsection{Prior Generation Strategy}
We selected 3 schemes for generating the prior dataset for adsorption isotherms. The first two were based of Latin hypercube sampling (LHS) in the feature space.\cite{LHS} For the first LHS-based sampling, pressure was sampled linearly from the pressure range. For the second LHS-based prior, pressure was sampled in a log (base 10) scale in the respective pressure range. Temperature was fixed at 300 K for section 3.1 and 3.2. For performing AL with 2 features (section 3.3 and 3.4), temperature was sampled linearly for both the LHS-based priors. 
In the third prior, named 'boundary-informed prior', samples were hand-picked at the limits of the test range (for two features, this would be a meshgrid of pressure and temperature points). For example, in case of methane adsorption with two features (section 3.3) we choose the pressure and temperature points as shown in table \ref{tbl:Prior}. The boundary-informed prior thus has 50 points for this section (five temperature points for each of the ten pressure mark). For section 3.4 (carbon dioxide adsorption with two features), boundary-informed prior had 40 points (8 pressure points with five temperature for each pressure mark). The pressure points of 200 and 300 bar (from table \ref{tbl:Prior}) were missing for carbon dioxide since the high pressure limit is 100 bar.\\

\begin{table}[ht]
\small
  \caption{\ Boundary-informed prior grid points for CH\textsubscript{4} adsorption in Cu-BTC MOF for two features}
  \label{tbl:Prior}
  \begin{tabular*}{0.48\textwidth}{@{\extracolsep{\fill}}cc}
    \hline
    Pressure (in bar) & Temperature (in K)\\
    \hline
    10\textsuperscript{-6} \\
    10\textsuperscript{-5} & 100\\
    10\textsuperscript{-4} \\
    10\textsuperscript{-3} & 150\\
    10\textsuperscript{-2} \\
    10\textsuperscript{-1} & 200\\
    1 \\
    10 & 250\\
    100 \\
    200 & 300\\ 
    300 \\
    \hline
  \end{tabular*}
\end{table}

\subsection{Error Calculation}

Three error metrics are used in the AL framework:\\
1. GP-predicted Relative Error --- This is the ratio of GP-predicted uncertainty at a point by the GP-predicted adsorption. Please note the aim of the AL procedure is to constrain the GP-predicted relative error within a threshold limit (refer to figure \ref{fgr:al_workflow} and \ref{fgr:methane Panel}).
\begin{equation}
    \text{GP relative Error in \% (at a point)} = \frac{{\sigma}_{\text{GP-predict}}(x)}{Y_{\text{GP-predict}}(x)} \times 100 
\end{equation}
2. Relative Error --- This is ratio of the difference between GP-predicted adsorption and the ground truth adsorption calculated by GCMC simulation.
\begin{equation}
    \text{Relative Error in \% (at a point)} = \Bigg|\frac{Y_{\text{GP-predict}}(x) - Y_{\text{GCMC}}(x)}{Y_{\text{GCMC}}(x)}\Bigg| \times 100 
\end{equation}
3. Mean Relative Error (MRE) --- This is calculated as a mean of the relative error for an entire AL iteration. We compare this error with the maximum GP relative error to check for speed of convergence of the AL protocol. Also, since MRE compares GP-predicted adsorption and ground truth based off of GCMC simulations, it serves as a parameter to gauge the performance of the AL model.

\begin{equation}
    \text{MRE in \%} = \left(\sum_{i=1}^{n} \Bigg|{\frac{Y_{\text{GP-predict}}(x_i) - Y_{\text{GCMC}}(x_i)}{Y_{\text{GCMC}}(x_i)}}\Bigg|\right) \times \frac{100}{n}
\end{equation}

\section{Results and Discussions}

\subsection{Methane Isotherms}
\begin{figure*}
\centering
  \includegraphics[height=9.5cm]{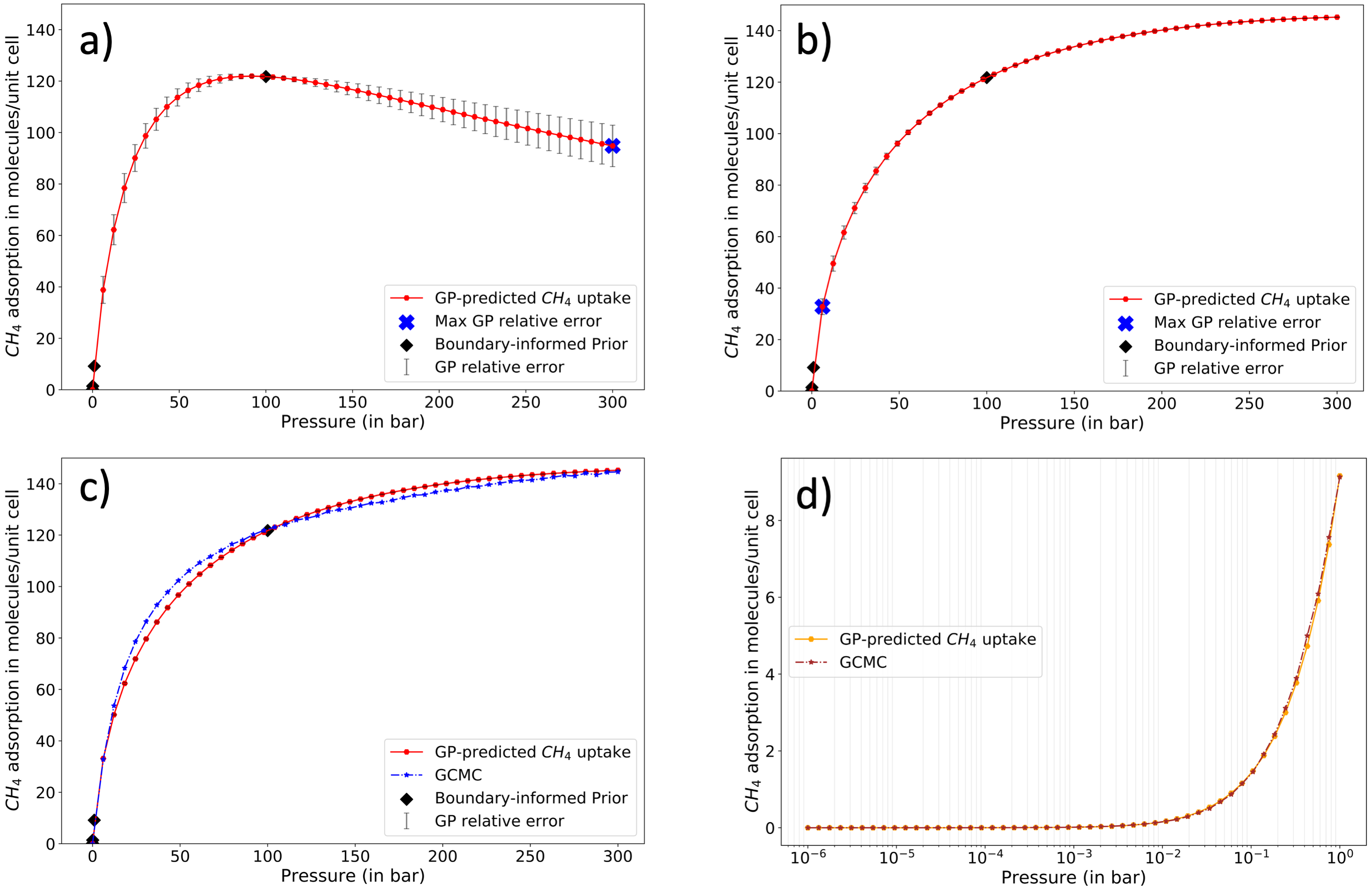}
  \caption{Evolution of GP fit to GCMC simulations of a methane isotherm at 300 K. Line in red represents the GP fit. Panel a shows the GP fit to the data resulting from boundary-informed prior selection.  Boundary-informed prior points are shown in black diamonds. Panels a-b show the subsequent iterations and the new simulation added to the data to be fit is shown in blue marker. Panel c shows the final GP fit along with results from GCMC simulations (ground truth) for a full methane isotherm. Panel d shows the GP fit along with GCMC simulations for the pressure range of 10\textsuperscript{-6} to 1 bar}
  \label{fgr:methane Panel}
\end{figure*}

Methane adsorption is Type I and is relatively simple to model as a single sphere without electrostatic interactions. Figure \ref{fgr:methane Panel} shows the evolution of the GP fit through all the iterations of the AL procedure for a methane isotherm at 300 K. Starting from 8 data points selected using boundary-informed prior scheme, only 2 iterations are needed to decrease the relative error of the GP fit (equation 4) to under 2 \%. For the LHS-based priors, four points were chosen for building the prior dataset. For boundary-informed prior, we find a good agreement between the GP predictions and the simulation results, and we show this case in figure \ref{fgr:methane Panel}. Panel a (in figure \ref{fgr:methane Panel}) shows the GP fit using the simulation data selected through boundary-informed prior. The first GP fit clearly struggles at high pressures where it under predicts methane loading and this is also reflected in high GP relative error (shown as grey bars in the plot). The highest relative standard deviation was at 300 bar and panel b shows the resulting GP fit with the new simulation result added to the data (blue marker). The GP fit now resembles more of what is expected of an adsorption isotherm.

\begin{figure*}
\centering
  \includegraphics[height=6.0cm]{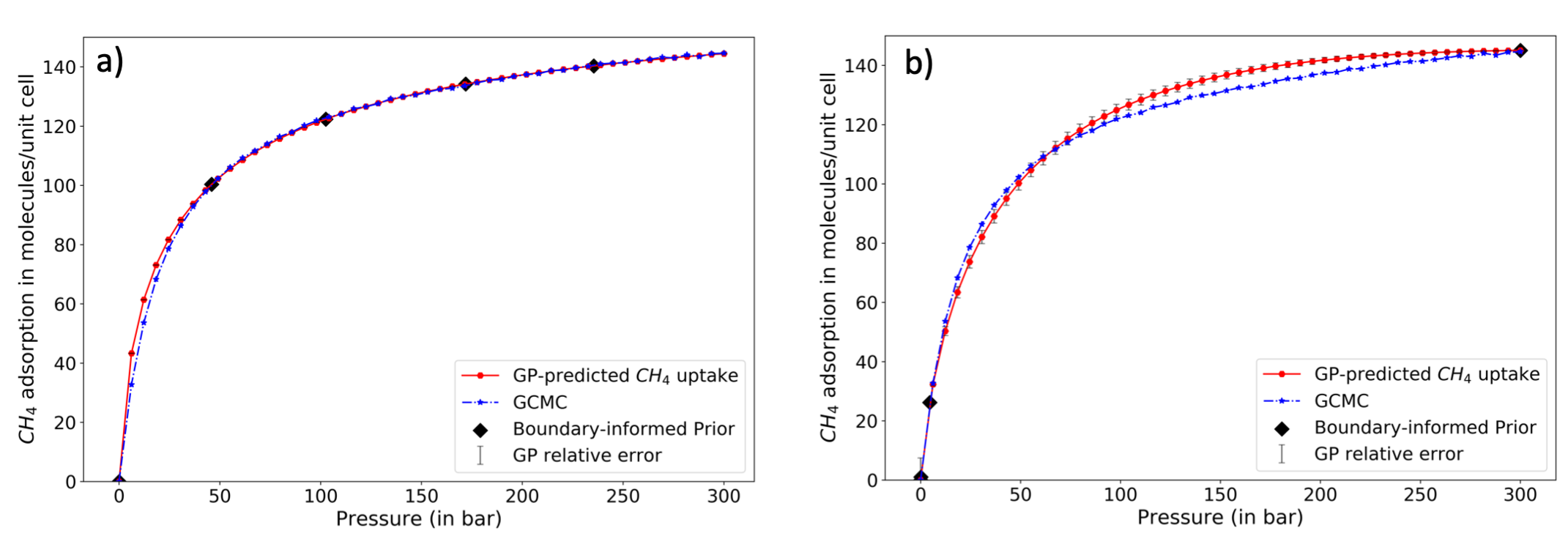}
  \caption{Final GP fit methane isotherm comparison with GCMC simulations (ground-truth) for a) linear-spaced LHS prior, and b) log-spaced LHS prior. We find the linear spaced prior has more deviations in the low-pressure region and has good accuracy for high pressure. For log-spaced prior there is large deviation at high pressure and moderate agreement with the GCMC simulation at low pressure}
  \label{fgr:methane_other_priors}
\end{figure*}

Qualitatively the fit does not change very much after the first iteration. Panel c shows the final GP fit compared to a full simulated isotherm with a good agreement between the GP fit and the GCMC simulation results. The last panel d shows the comparison of GP fit with GCMC simulations at low pressure range (10\textsuperscript{-6} to 1 bar). The final adsorption isotherm GP fits along with the GCMC simulations for both linear-spaced and log-spaced LHS have been included in the Supporting Information (figure \ref{fgr:methane_other_priors}). 

The performance of other priors, linear-LHS and log-LHS, along with boundary-informed are tabulated in table \ref{tbl:methane_1f}. As we can observe, the overall MRE (for ($X_{test}$)) is only 1.15 \% for linear-spaced prior while it is slightly higher for boundary-informed prior at 2.14 \%. The log-spaced prior has a very high MRE of 7.80 \% for $X_{test}$ compared to the other priors. For the low pressure interpolation test, the log-spaced LHS finished after one iteration and performed the best among all the prior with an MRE of 13.61 \%. However, the MRE of the log-based prior was high for the $X_{test}$ range. Overall the GP fits obtained using the previous protocols shows poor performance in the low pressure regime (10\textsuperscript{-6} to 1 bar) for all the priors, especially for linear-spaced one. In some cases the predicted adsorption can differ by an order of magnitude. Hence, despite the perceived agreement of the GP fit, the errors in the low pressure region are high when comparing with simulation results. This can present significant challenges in analysis for separations where ideal adsorbed solution theory (IAST) is used and it is particularly sensitive to the results in the low pressure regime.\cite{IAST_50}

\begin{table}[ht]
\small
  \caption{\ Performance of different priors for predicting CH\textsubscript{4} uptake in Cu-BTC MOF (all errors are expressed in \%)} 
  \label{tbl:methane_1f}
  \begin{tabular*}{0.48\textwidth}{@{\extracolsep{\fill}}cccc}
    \hline
   Prior type & Iterations & MRE ($X_{test}$) & MRE (Low Pressure)\\
    \hline
    Boundary-informed & 2 & 2.14 & 15.79\\
    Linear-spaced LHS & 1 & 1.15 & 37.84\\
    Log-spaced LHS & 1 & 7.80 & 13.61\\
    \hline
    GCMC & - & $\sigma_{GCMC}$ ($X_{test}$) & $\sigma_{GCMC}$ (Low Pressure)\\ 
    \hline
    Ground truth & - & 0.74 & 73.97\\
    \hline
  \end{tabular*}
\end{table}
These three AL approaches for adsorption isotherms show good agreement with the GCMC simulations despite not having to physically simulate all the points in the test set. The first approach only used eight data points for the prior, including one at each boundary of the isotherm in pressure, and the AL procedure converges with only two additional iterations. For the linear-spaced LHS approach, the low pressure regimes of the isotherm are problematic and resulted in a high MRE. The log-spaced prior, shows a better MRE at low pressure than the boundary-informed approach, but its performance was poor for pressure range in the $X_{test}$. The code for the AL along with the data are publicly available and can be accessed through this link: \href{https://github.com/mukherjee07/Sequential_Design_Adsorption_small_molecules_in_MOFs}{https://github.com/mukherjee07/Sequential\textunderscore Design\textunderscore of\textunderscore\newline Adsorption\textunderscore simulations\textunderscore in\textunderscore MOFs}.

\subsection{Carbon Dioxide Isotherms}

Carbon dioxide adsorption isotherms present an interesting contrast to methane adsorption. Namely, the electrostatic interactions of the molecule induce a much sharper transition in the isotherm. Despite this, we see very similar behavior and results for the AL procedures for carbon dioxide when compared to methane. We carried out AL with three priors, as was done for methane. The first difference in this study was the total pressure range, which was 10\textsuperscript{-6} to 100 bar. Another difference is 9 points were chosen for the boundary-informed prior since the transition is sharper for carbon dioxide adsorption. For each of the log-spaced and linear-spaced priors, four points in the isotherm were generated in an automated fashion similar to methane adsorption.

The AL converged to a 2 \% GP relative error in a similar number of iterations as for methane, except for log-spaced prior where it took 6 iterations. Table \ref{tbl:CO2_1f} shows the final GP fit results compared to the simulated isotherm. Boundary-informed prior GP fit had the best MRE at both high pressure ($X_{test}$) and low pressure range.
\begin{table}[ht]
\small
  \caption{\ Performance of different priors for predicting CO\textsubscript{2} uptake in Cu-BTC MOF (all errors are expressed in \%)} 
  \label{tbl:CO2_1f}
  \begin{tabular*}{0.48\textwidth}{@{\extracolsep{\fill}}cccc}
    \hline
   Prior type & Iterations & MRE ($X_{test}$) & MRE (Low Pressure)\\
    \hline
    Boundary-informed & 3 & 1.52 & 20.39\\
    Linear-spaced LHS & 1 & 3.66 & 1011.21\\
    Log-spaced LHS & 6 & 2.07 & 73.37\\
    \hline
    GCMC & - & $\sigma_{GCMC}$ ($X_{test}$) & $\sigma_{GCMC}$ (Low Pressure)\\ 
    \hline
    Ground truth & - & 1.04 & 51.95\\
    \hline
  \end{tabular*}
\end{table}

\begin{figure*}
\centering
  \includegraphics[height=12.0cm]{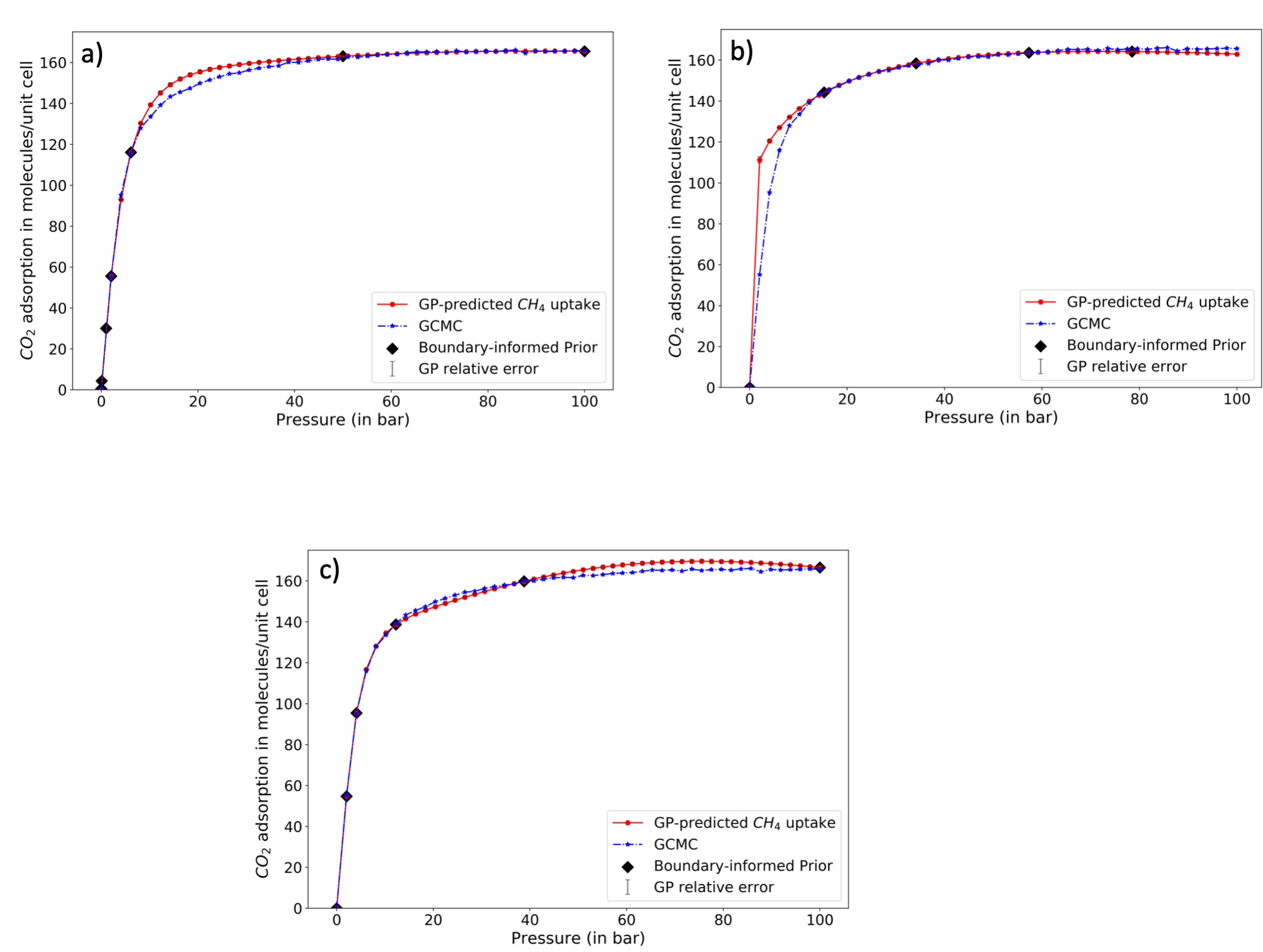}
  \caption{Final GP fit CO\textsubscript{2} isotherm comparison with GCMC simulations (ground-truth) for a) boundary-informed prior, b) linear-spaced LHS prior, and c) log-spaced LHS prior. We observe that boundary-informed prior overpredicts at the adsorption rise zone but performs very well at both the low- and high-pressure region. For the LHS-based priors, as was observed for methane adsorption, linear-spaced prior performs better at high pressure while log-spaced one does significantly better at low pressure}
  \label{fgr:CO2_1f_all}
\end{figure*}

The GP fit carbon dioxide isotherm are shown in the Supporting Information figure \ref{fgr:CO2_1f_all}. We find the final GP fit for boundary-informed prior performs excellently in the $X_{test}$ pressure range as well in the low pressure region. The log-spaced one performs well at the high pressure region ($X_{test}$) while the linear-spaced one has high error at low pressure as well as at the tail of the pressure range. This also becomes evident from the MRE for both the LHS-based prior schemes in table \ref{tbl:CO2_1f}. The linear-spaced prior based GP fit has a MRE of 3.66 \% for the $X_{test}$ isotherm with a very high MRE of 1011.21 \% for the low pressure range. The log-spaced GP fit performs better than the linear one but still has a higher MRE compared to boundary-informed for both the pressure ranges. Through this comparison, it is evident that boundary-informed prior outperforms both LHS schemes when using pressure as a single feature. 

\subsection{Temperature-Pressure diagrams for methane adsorption}

We performed adsorption simulation in the temperature and pressure phase-space (two features for AL) with priors based on boundary-informed and LHS sampling schemes. The pressure and temperature range for this study was 10\textsuperscript{-6} bar to 300 bar, and 100 K to 300 K respectively. The boundary-informed one, similar to the previous methane and carbon dioxide isotherms, was curated to bias the training data with hand-picked pressure and temperature points. For each pressure point as reported in table \ref{tbl:Prior} five temperature points (100 K, 150 K, 200 K, 250 K, and 300 K) were chosen to form a 50 point prior. The other priors were LHS based, linear and log-spaced, sampled along the temperature and pressure phase space. Both the LHS and log-based prior had also 50 points for a fair comparison with the boundary-informed prior.

The ground truth dataset was created using GCMC simulations for two separate tests as explained previously. Like, adsorption for a single feature, $X_{test}$ was linearly spaced with 50 points between 10\textsuperscript{-6} to 300 bar, which was biased for the high pressure region. This same dataset had 40 temperature points divided linearly from 100 K to 300 K for each pressure. Thus $X_{test}$ had 2000 points. For the low-pressure interpolation ground truth, the pressure range was from 10\textsuperscript{-6} to 1 bar, with 50 pressure points linearly distributed in the log-space of this range. The temperature points was distributed linearly as $X_{test}$, with 40 points in temperature for each pressure. Thus the low-pressure ground truth also had 2000 points. The AL fit was done with $X_{test}$, and so the AL protocol had zero knowledge of the adsorption in low-pressure region. This was purposefully done to test the power of the method for interpolating to low-pressure region similar to the one for single feature in section 3.1 and 3.2.

\begin{figure}[ht]
\centering
  \includegraphics[height=6.5cm]{Boundary-informed-prior-CH4.png}
  \caption{Comparison of GP-predicted CH\textsubscript{4} uptake with GCMC simulation predicted for pressure range from 10\textsuperscript{-6} to 300 bar, at temperature of 100 K, 202 K and 300 K for boundary-informed prior.}
  \label{fgr:boundary_ch4_uptake}
\end{figure}

The best performing prior for this study was boundary-informed and the final GP fit with GCMC simulation is shown in figure \ref{fgr:boundary_ch4_uptake}. 
The GP fit predicts the uptake very close to the GCMC and the MRE is only 0.86 \% for $X_{test}$ as reported in table \ref{tbl:CH4_2f}. 
With only a total of 33 iterations, it can predict the uptake for a phase space of 2000 points with a very low MRE (less than 1 \%). The log-spaced prior based GP fit had a slightly higher MRE of 7.99 \% followed by linearly-spaced GP fit with 8.62 \%. The number of iterations for the log-spaced was 19, comparable with that of boundary-informed while for the linear-spaced it was only 6.

\begin{figure*}
\centering
  \includegraphics[height=4.0cm]{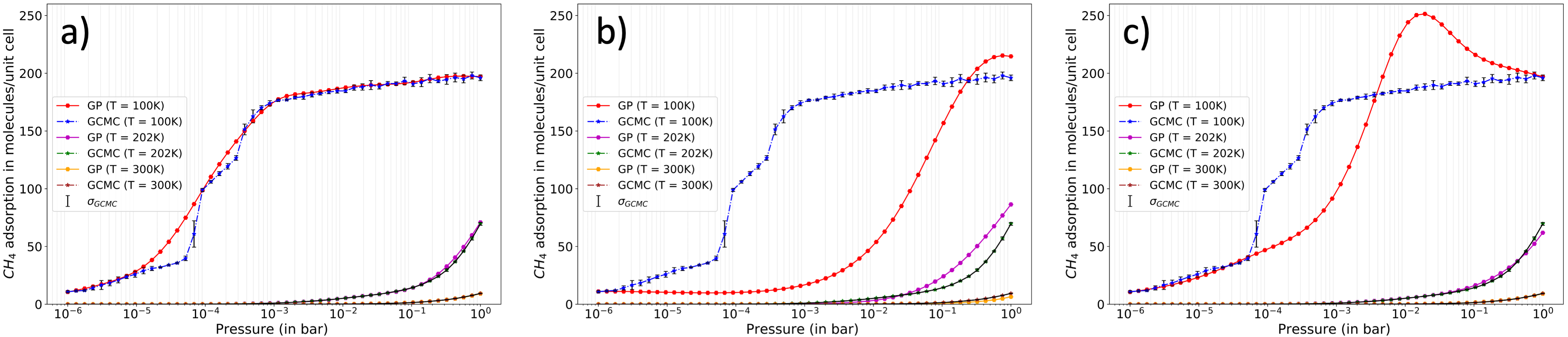}
  \caption{Methane uptake comparison between GP and GCMC simulation in Cu-BTC at low pressure range for different priors, a) Boundary-informed prior, b) Linear-spaced LHS prior, and c) Log-spaced LHS prior}
  \label{fgr:methane_2F}
\end{figure*}

Though each of these priors performed reasonably well in the $X_{test}$ range, the low-pressure test set revealed appreciable differences in their performance. Observing the MRE for low pressure interpolation, we find that log-spaced prior is at 13.43 \%, which is the best among the three. This was followed by boundary-informed at 18.30 \%, and then we had the linear-spaced LHS with a very poor MRE of 85.74 \%. The methane uptake at low pressure indicates that the performance of these models are comparable. However observing the methane uptake at this range, as shown in figure \ref{fgr:methane_2F}, we find a substantial difference in their predictions especially at the lowest temperature of 100 K. In figure \ref{fgr:methane_2F}a, the boundary-informed prior based uptake performs reasonably well at 100 K while in \ref{fgr:methane_2F}b and \ref{fgr:methane_2F}c, the GP fits from linear and log-spaced LHS prior are very far off from ground truth. The log-spaced prior first over predicts then returns to the GCMC simulation range while the linear-based prior under predicts the GCMC ground truth. The situation improves for both the LHS schemes at higher temperature of 202 K and 300 K since here the GP fit starts to match the ground truth. 

\begin{figure*}
\centering
  \includegraphics[height=3.85cm]{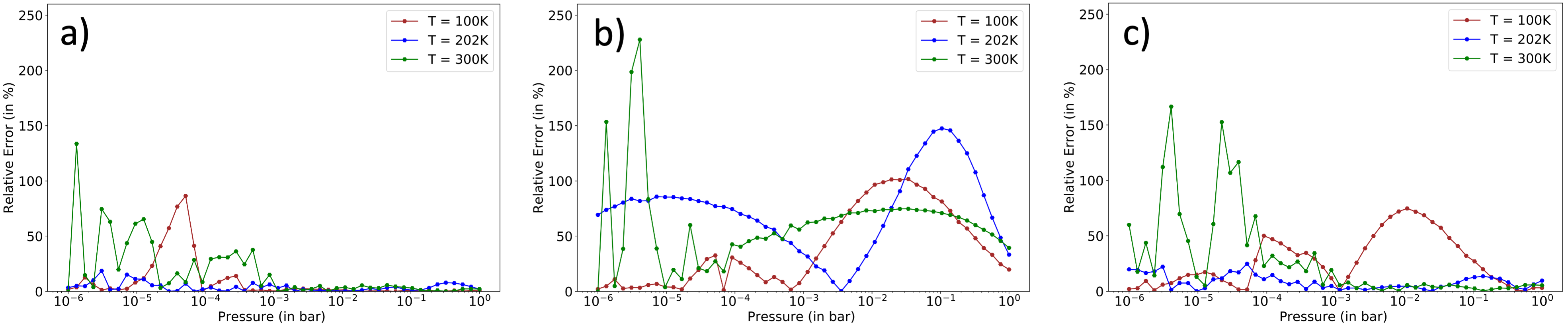}
  \caption{Relative error (in \%) comparison between GP and GCMC simulation in Cu-BTC at low pressure range for different priors, a) Boundary-informed prior, b) Linear-spaced LHS prior, and c) Log-spaced LHS prior}
  \label{fgr:RE_2F_methane}
\end{figure*}

\begin{figure*}
\centering
  \includegraphics[height=6.0cm]{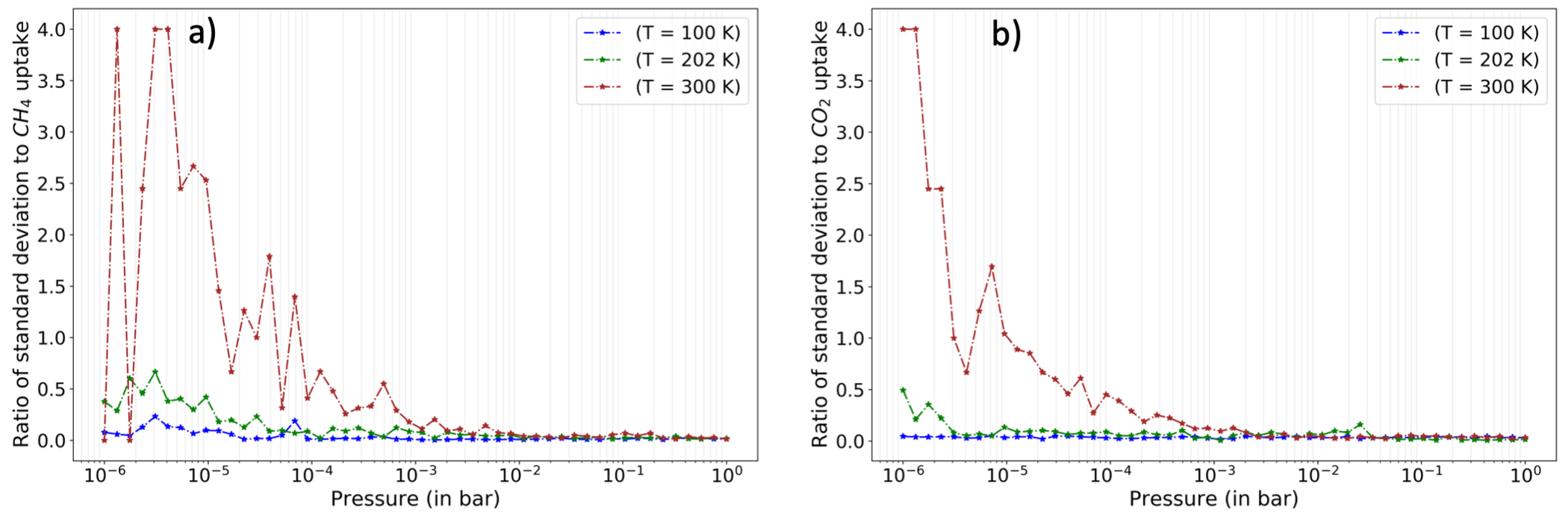}
  \caption{Ratio of standard deviation in GCMC simulation to gas uptake at low pressure range (10\textsuperscript{–6} to 1 bar) in Cu-BTC at different temperatures, a) Methane adsorption, and b) CO\textsubscript{2} adsorption. Adsorption is extremely low (almost nil) for both these gases at high temperature of 300 K, and hence this ratio is very high, especially for the 10\textsuperscript{–6} to 10\textsuperscript{–4} bar range. Since the absolute adsorption is near zero, it is extremely difficult to predict the uptake accurately at this region. Even for temperature of 202 K we see this ratio remains well above 0.5 for a considerable range of pressure. These plots illustrate the high degree of uncertainty in the GCMC simulations and hence the ground truth has a high unreliability in this space which also is manifested in the final GP fit}
  \label{fgr:std_y_GCMC_lowP}
\end{figure*}

This behaviour is also reflected in relative error isotherm plot at the low pressure range in figure \ref{fgr:RE_2F_methane}. We see for boundary-informed prior, the highest relative error is 140 \% at a single point and the rest of the errors are less than 100 \% at these temperatures. The relative errors are comparatively very high for linear-spaced and log-spaced prior schemes. Though it can be pointed out that error range of 50 \% for the boundary-informed prior is still high for predictions, we should observe that the uncertainty of the GCMC simulations is also very high in this range (table \ref{tbl:CH4_2f}). In figure \ref{fgr:std_y_GCMC_lowP} in the Supporting Information, we have shown the ratio of standard deviation to methane uptake at low pressure for the GCMC simulation, and we can find that this ratio is well above 1.0 (more than 100 \% error) for a major portion of this space and in the extremely low-pressure region it is as high as 4.0 (this is true for pressure range 10\textsuperscript{-6} to 10\textsuperscript{-4} bar). Given the high uncertainty in GCMC simulation in this region, the boundary-informed prior uptake results can be accepted. Also, in figure \ref{fgr:RE_2F_methane} and \ref{fgr:methane_2F}, we find that boundary-informed prior has low relative error than log-spaced prior but in table \ref{tbl:CH4_2f}, we see the MRE of log-spaced prior is lower than boundary-informed. This can be explained from our choice of temperature, which were 100 K, 202 K and 300 K, for figures \ref{fgr:methane_2F}, and \ref{fgr:RE_2F_methane}, points which corresponded to the prior points in boundary-informed. The log-spaced prior was sampled in LHS and hence the temperature had a wide distribution and hence log-spaced prior would overall outperform boundary-informed prior if we take the complete temperature range into consideration. However, at the temperature boundaries (figures \ref{fgr:methane_2F}, and \ref{fgr:RE_2F_methane}), the boundary-informed prior would have lower errors than log-spaced ones.

\begin{figure}[ht]
\centering
  \includegraphics[height=6cm]{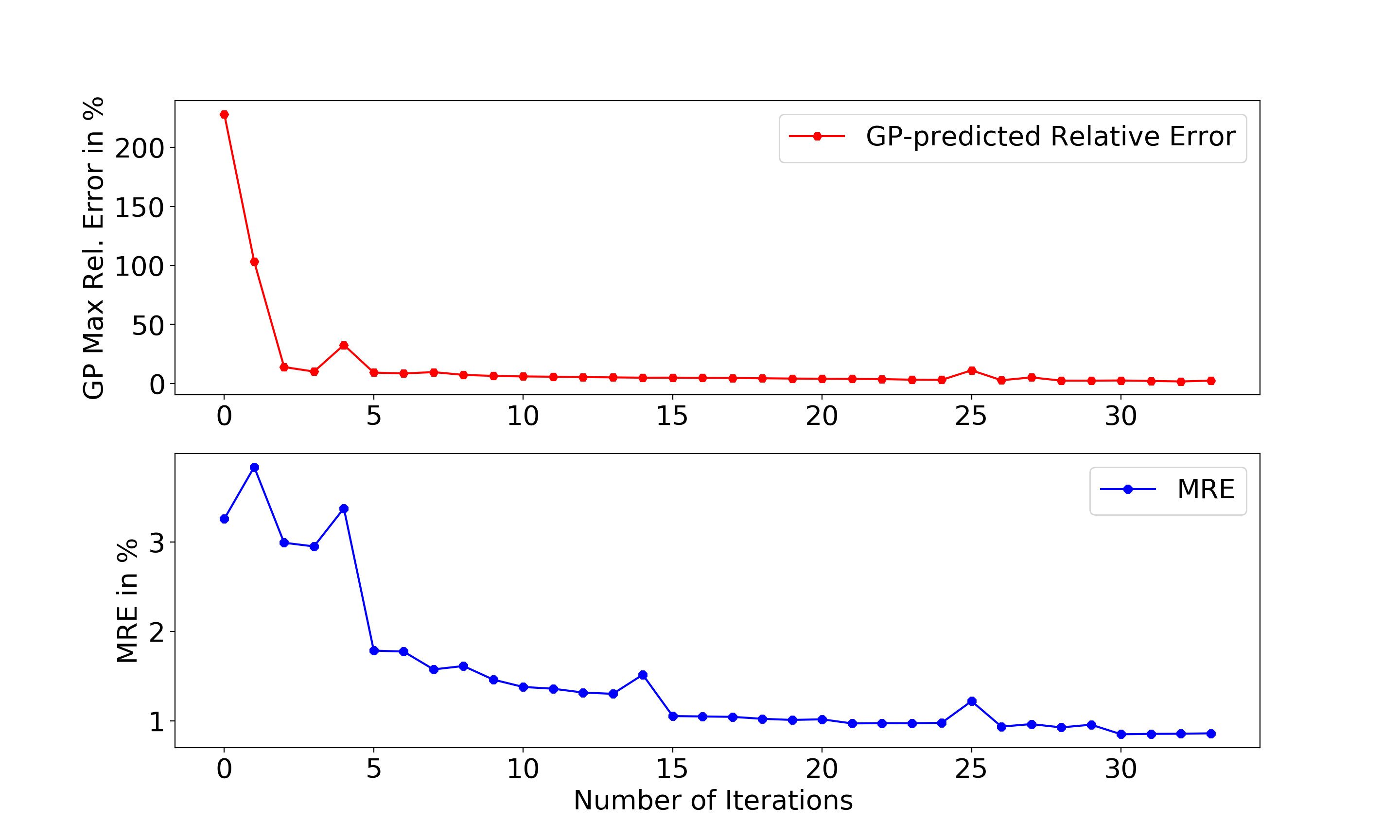}
  \caption{Maximum GP-predicted relative error and MRE (mean relative error) with AL for methane adsorption in Cu-BTC with iterations for boundary-informed prior.}
  \label{fgr:boundary-rel-rrmse-methane}
\end{figure}

\begin{table}[ht]
\small
  \caption{\ Performance of different priors for predicting CH\textsubscript{4} uptake in Cu-BTC MOF with two features (all errors are expressed in \%)} 
  \label{tbl:CH4_2f}
  \begin{tabular*}{0.48\textwidth}{@{\extracolsep{\fill}}cccc}
    \hline
   Prior type & Iterations & MRE ($X_{test}$) & MRE (Low Pressure)\\
    \hline
    Boundary-informed & 33 & 0.86 & 18.30\\
    Linear-spaced LHS & 6 & 8.62 & 85.74\\
    Log-spaced LHS & 19 & 7.99 & 13.43\\
    \hline
    GCMC & - & $\sigma_{GCMC}$ ($X_{test}$) & $\sigma_{GCMC}$ (Low Pressure)\\ 
    \hline
    Ground truth & - & 3.18 & 23.53\\
    \hline
  \end{tabular*}
\end{table}

 
Another aspect of this study is the convergence of AL with iterations. Figure \ref{fgr:boundary-rel-rrmse-methane} presents AL based on boundary-informed prior convergence in terms of maximum GP-predicted relative error and MRE. Since the AL continues until the maximum GP relative error is less than 2 \%, it takes a number of iterations before the protocol converges. In figure \ref{fgr:boundary-rel-rrmse-methane} we can observe that the GP maximum error quickly goes to a very low point (say 3 \%). However to reach 2 \% maximum error for the GP, it takes a large number of iterations. For boundary-informed prior it took 33 iterations to converge. While 33 iterations of AL was quite fast for methane adsorption, a molecule which doesn't have electrostatic interactions, this aspect can play an important role for complex molecules. We will address this issue further for carbon dioxide adsorption in the next section and examine how fast the boundary-informed prior errors are converging with respect to iterations.


\subsection{Temperature-Pressure diagrams for carbon dioxide adsorption}
As mentioned earlier, carbon dioxide adsorption on Cu-BTC is more complex than methane adsorption due to electrostatic interactions. For carbon dioxide adsorption, the boundary-informed prior performs the best. However, AL converges very slowly for carbon dioxide and hence for this case, we changed the limit of maximum GP relative error (which was 2 \% for all cases before) convergence limit to 3 \%. In table \ref{tbl:CO2_two_features_Prior}, the MRE reported were based on prior convergence of maximum GP relative error of 3 \%. One interesting observation is that boundary-informed MRE at low pressure for carbon dioxide adsorption with a 3 \% cut-off is closer to that of methane at the threshold of 2 \%. This might be due to a high value of maximum uncertainty in the low pressure region for the case of carbon dioxide adsorption, and so to obtain a flat GP relative error, AL needs more iterations. However, since MRE presents a mean property of the relative error, the majority of the points for carbon dioxide adsorption had a lower error for this low pressure and hence the MRE was also smaller. We also observed that the linear-prior and log-prior took more iterations, 10 and 50 respectively, in case of carbon dioxide to get a maximum GP relative error of 3 \%, than methane, which was only 6 and 19 to a achieve a 2 \% maximum GP relative error. 

In figure \ref{fgr:boundary_co2_uptake}, we have shown the final GP fit based on boundary-informed prior compared with GCMC simulations (ground truth). We find a very close agreement between the GP fit and GCMC calculations. The uncertainty (shown as $\sigma_{GCMC}$), however, is very high for temperature of 100 K and here the GP under predicts the carbon dioxide uptake in the mid-pressure range. However this error is very small and is close to the 2 \% relative error limit.
\begin{figure}[ht]
\centering
  \includegraphics[height=6.5cm]{Boundary-informed-prior-CO2.png}
  \caption{Comparison of GP-predicted CO\textsubscript{2} uptake with GCMC simulation predicted for pressure range from 10\textsuperscript{-6} to 100 bar, at temperature of 100 K, 202 K and 300 K for boundary-informed prior.}
  \label{fgr:boundary_co2_uptake}
\end{figure}

\begin{figure}[ht]
\centering
  \includegraphics[height=6.0cm]{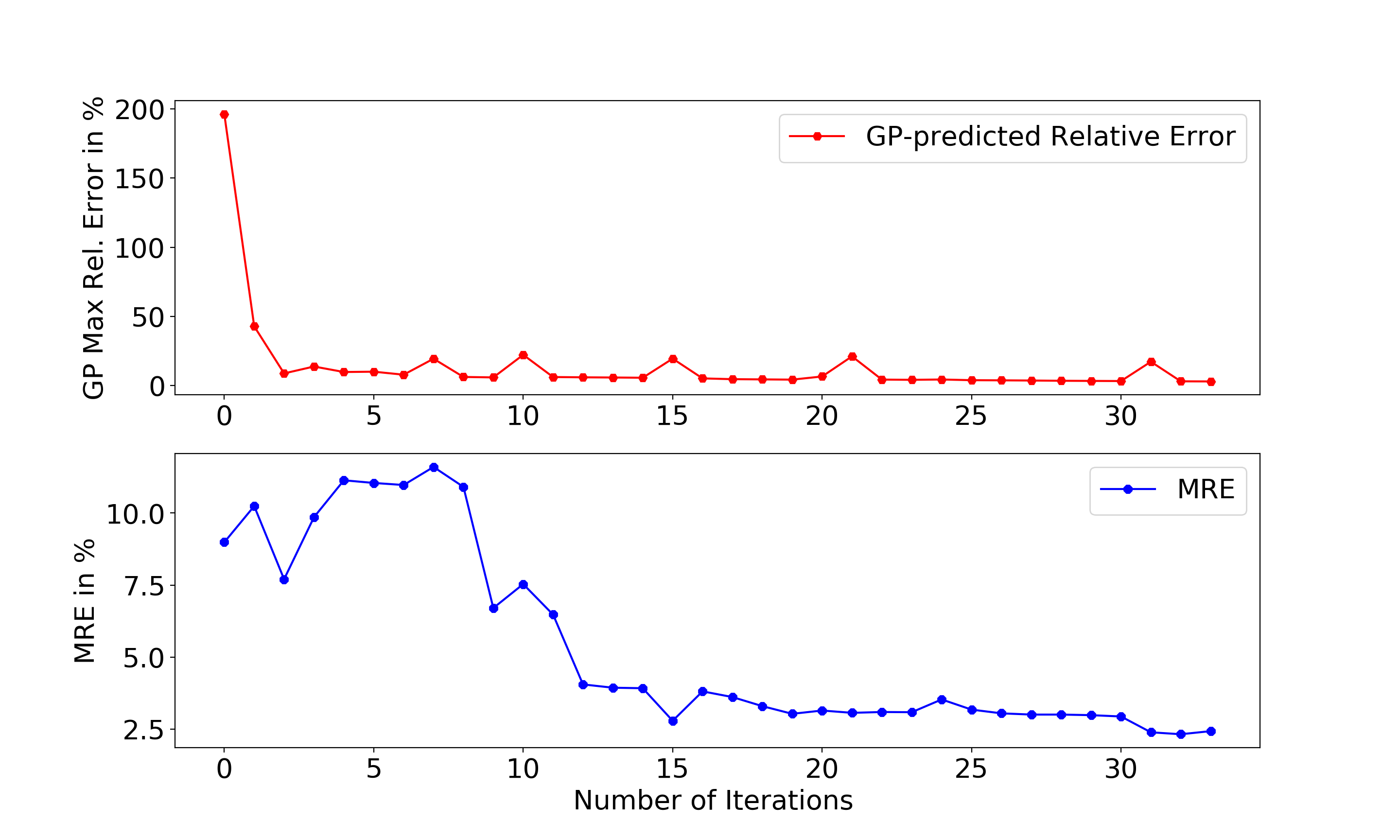}
  \caption{Maximum GP-predicted relative error and MRE (mean relative error) with AL for CO\textsubscript{2} adsorption in Cu-BTC with iterations for boundary-informed prior.}
  \label{fgr:boundary-rel-rrmse-CO2}
\end{figure}

As discussed before, convergence of maximum GP relative error with iterations is very slow for carbon dioxide (shown in figure \ref{fgr:boundary-rel-rrmse-CO2} for the boundary-informed prior). It took 33 iterations for the maximum GP relative error to reach 3 \%, however it takes 129 iterations to reach the limit of 2 \%. Still, the performance of a 3 \% convergence is very good and comparable to methane adsorption at the 2 \% threshold. The MRE, as shown in table \ref{tbl:CO2_two_features_Prior} at both the full pressure and low pressure ranges are comparable, if not lower, than that of methane. 

Apart from the slow convergence and encountering higher uncertainties at low temperature, AL does manage to predict carbon dioxide uptake with comparable accuracy with that of methane for two features. This further proves that the method is transferable to complex molecules and we can also effectively explore the adsorption conditions of temperature and pressure (including the low pressure region) for these complex molecules with a limited number of simulations dictated by AL.
\begin{table}[ht]
\small
  \caption{\ Performance of different priors for predicting CO\textsubscript{2} uptake in Cu-BTC MOF with two features (all errors are expressed in \%)} 
  \label{tbl:CO2_two_features_Prior}
  \begin{tabular*}{0.48\textwidth}{@{\extracolsep{\fill}}cccc}
    \hline
    Prior type & Iterations & MRE ($X_{test}$) & MRE (Low Pressure)\\
    \hline
    Boundary-informed & 33 & 2.43 & 18.10\\
    Linear-spaced LHS & 9 & 2.79 & 43.11\\
    Log-spaced LHS & 49 & 2.64 & 18.07\\
    \hline
    GCMC & - & $\sigma_{GCMC}$ ($X_{test}$) & $\sigma_{GCMC}$ (Low Pressure)\\ 
    \hline
    Ground truth & - & 3.47 & 15.64\\
    \hline
  \end{tabular*}
\end{table}

\section{Conclusions}
Based on the methane and carbon dioxide adsorption proof-of-concept case studies, we can conclude that the AL framework is a promising method to efficiently collect data from molecular simulations, and the trained GPR surrogate models can replace GCMC simulation for emulating adsorption isotherm. For the case of pressure and temperature adsorption space for methane and carbon dioxide (section 3.3 and 3.4), we showed that with only 33 iterations of AL iterations, the algorithm can predict 4000 data points in temperature and the pressure range. This includes the low pressure region which is important for separation predictions (IAST). We can recognize here that with less than 2 \% of the data AL can accurately estimate the full isotherms for a large temperature and pressure range. Having a protocol like AL to sequentially select adsorption simulations for surrogate models can save orders of magnitude in terms of computational cost in designing cheap and reliable surrogate model for adsorption prediction.

AL is also much faster than GCMC simulations and a GPR surrogate model only takes a few seconds to a few minutes to predict the whole isotherm. If we take the complete pressure and temperature space, the computational cost of the GPR remains very low, and the prediction is finished within minutes. However a single GCMC adsorption simulation at a fixed pressure and temperature can take from a few minutes to a few hours (can also go beyond a day depending on molecule complexity and number of production runs). Thus, predicting a full isotherm (with 50 points) can take a day or longer for complex molecules, while performing a pressure-temperature phase space simulation can take between a week to a month in terms of computational cost. In essence, AL is order of magnitudes faster than conventional GCMC simulation for predicting adsorption simulation in MOFs.

Among the priors we tested, the boundary-informed one performed best considering both the $X_{test}$ and low pressure interpolation dataset. We also found the log-spaced LHS prior can outperform boundary-informed prior in the low pressure range but has large relative errors at the high pressure region. Similarly, the linear-spaced LHS prior generally performs well at high pressures but is very poor in the low pressure range. In contrast, the boundary-informed prior has a good balance of both the low and high pressure points, and thus comes across as a better choice for building priors for AL. In this context more novel prior models can be explored including schemes like orthogonal arrays and composite designs.\cite{novel_priors}

Alternative AL approaches can also be explored, including the addition of multiple sampling points in a parallel fashion during the building of the GP model. In each iteration, we can select multiple points for sampling which have a GP predicted relative errors above a set uncertainty threshold. While, this study presents a simple application of AL for relatively simple molecules (methane and carbon dioxide), further studies on the number of features and other aspects of AL are needed to comprehensively understand the usefulness of AL for adsorption in MOFs.
\section*{Conflicts of interest}
There are no conflicts to declare.

\section*{Acknowledgements}
KM would like to thank the ND Energy's Eilers Family Graduate fellowship. YJC acknowledges support from the University of Notre Dame and AWD acknowledges support from the U.S.~National Science Foundation under award CBET-1941596. All authors thank the ND Center for Research Computing for computation resources and technical support.


\balance


\bibliography{rsc} 

\providecommand*{\mcitethebibliography}{\thebibliography}
\csname @ifundefined\endcsname{endmcitethebibliography}
{\let\endmcitethebibliography\endthebibliography}{}
\begin{mcitethebibliography}{59}
\providecommand*{\natexlab}[1]{#1}
\providecommand*{\mciteSetBstSublistMode}[1]{}
\providecommand*{\mciteSetBstMaxWidthForm}[2]{}
\providecommand*{\mciteBstWouldAddEndPuncttrue}
  {\def\EndOfBibitem{\unskip.}}
\providecommand*{\mciteBstWouldAddEndPunctfalse}
  {\let\EndOfBibitem\relax}
\providecommand*{\mciteSetBstMidEndSepPunct}[3]{}
\providecommand*{\mciteSetBstSublistLabelBeginEnd}[3]{}
\providecommand*{\EndOfBibitem}{}
\mciteSetBstSublistMode{f}
\mciteSetBstMaxWidthForm{subitem}
{(\emph{\alph{mcitesubitemcount}})}
\mciteSetBstSublistLabelBeginEnd{\mcitemaxwidthsubitemform\space}
{\relax}{\relax}

\bibitem[Kondo \emph{et~al.}(1997)Kondo, Yoshitomi, Matsuzaka, Kitagawa, and
  Seki]{Kondo1997}
M.~Kondo, T.~Yoshitomi, H.~Matsuzaka, S.~Kitagawa and K.~Seki, \emph{Angewandte
  Chemie}, 1997, \textbf{36}, 1725--1727\relax
\mciteBstWouldAddEndPuncttrue
\mciteSetBstMidEndSepPunct{\mcitedefaultmidpunct}
{\mcitedefaultendpunct}{\mcitedefaultseppunct}\relax
\EndOfBibitem
\bibitem[Langmi \emph{et~al.}(2014)Langmi, Ren, North, Mathe, and
  Bessarabov]{Langmi2014}
H.~W. Langmi, J.~Ren, B.~North, M.~Mathe and D.~Bessarabov,
  \emph{Electrochimica Acta}, 2014, \textbf{128}, 368--392\relax
\mciteBstWouldAddEndPuncttrue
\mciteSetBstMidEndSepPunct{\mcitedefaultmidpunct}
{\mcitedefaultendpunct}{\mcitedefaultseppunct}\relax
\EndOfBibitem
\bibitem[Boyd \emph{et~al.}(2019)Boyd, Chidambaram, Garc{\'i}a-D{\'i}ez,
  Ireland, Daff, Bounds, G{\l}adysiak, Schouwink, Moosavi, Maroto-Valer,
  Reimer, Navarro, Woo, Garcia, Stylianou, and Smit]{CO2}
P.~Boyd, A.~Chidambaram, E.~Garc{\'i}a-D{\'i}ez, C.~Ireland, T.~Daff,
  R.~Bounds, A.~G{\l}adysiak, P.~Schouwink, S.~Moosavi, M.~Maroto-Valer,
  J.~Reimer, J.~Navarro, T.~Woo, S.~Garcia, K.~Stylianou and B.~Smit,
  \emph{Nature}, 2019, \textbf{576}, 253--256\relax
\mciteBstWouldAddEndPuncttrue
\mciteSetBstMidEndSepPunct{\mcitedefaultmidpunct}
{\mcitedefaultendpunct}{\mcitedefaultseppunct}\relax
\EndOfBibitem
\bibitem[Hu \emph{et~al.}(2019)Hu, Wang, Shah, and Zhao]{HU_CO2}
Z.~Hu, Y.~Wang, B.~B. Shah and D.~Zhao, \emph{Advanced Sustainable Systems},
  2019, \textbf{3}, 1800080\relax
\mciteBstWouldAddEndPuncttrue
\mciteSetBstMidEndSepPunct{\mcitedefaultmidpunct}
{\mcitedefaultendpunct}{\mcitedefaultseppunct}\relax
\EndOfBibitem
\bibitem[Pascanu \emph{et~al.}(2019)Pascanu, González~Miera, Inge, and
  Martín-Matute]{mof_catalysis}
V.~Pascanu, G.~González~Miera, A.~K. Inge and B.~Martín-Matute, \emph{Journal
  of the American Chemical Society}, 2019, \textbf{141}, 7223--7234\relax
\mciteBstWouldAddEndPuncttrue
\mciteSetBstMidEndSepPunct{\mcitedefaultmidpunct}
{\mcitedefaultendpunct}{\mcitedefaultseppunct}\relax
\EndOfBibitem
\bibitem[Lin \emph{et~al.}(2019)Lin, Xiang, Xing, Zhou, and Chen]{separation}
R.-B. Lin, S.~Xiang, H.~Xing, W.~Zhou and B.~Chen, \emph{Coordination Chemistry
  Reviews}, 2019,  87--103\relax
\mciteBstWouldAddEndPuncttrue
\mciteSetBstMidEndSepPunct{\mcitedefaultmidpunct}
{\mcitedefaultendpunct}{\mcitedefaultseppunct}\relax
\EndOfBibitem
\bibitem[Kreno \emph{et~al.}(2012)Kreno, Leong, Farha, Allendorf, Van~Duyne,
  and Hupp]{mof_sensors}
L.~E. Kreno, K.~Leong, O.~K. Farha, M.~Allendorf, R.~P. Van~Duyne and J.~T.
  Hupp, \emph{Chemical Reviews}, 2012, \textbf{112}, 1105--1125\relax
\mciteBstWouldAddEndPuncttrue
\mciteSetBstMidEndSepPunct{\mcitedefaultmidpunct}
{\mcitedefaultendpunct}{\mcitedefaultseppunct}\relax
\EndOfBibitem
\bibitem[Wang \emph{et~al.}(2018)Wang, Zheng, and Xie]{mof_drug}
L.~Wang, M.~Zheng and Z.~Xie, \emph{J. Mater. Chem. B}, 2018, \textbf{6},
  707--717\relax
\mciteBstWouldAddEndPuncttrue
\mciteSetBstMidEndSepPunct{\mcitedefaultmidpunct}
{\mcitedefaultendpunct}{\mcitedefaultseppunct}\relax
\EndOfBibitem
\bibitem[Wang \emph{et~al.}(2018)Wang, Zheng, and Xie]{mof_drugs}
L.~Wang, M.~Zheng and Z.~Xie, \emph{J. Mater. Chem. B}, 2018, \textbf{6},
  707--717\relax
\mciteBstWouldAddEndPuncttrue
\mciteSetBstMidEndSepPunct{\mcitedefaultmidpunct}
{\mcitedefaultendpunct}{\mcitedefaultseppunct}\relax
\EndOfBibitem
\bibitem[Fritz \emph{et~al.}(2021)Fritz, Colón, and Herrera]{photon}
R.~A. Fritz, Y.~J. Colón and F.~Herrera, \emph{Chem. Sci.}, 2021, \textbf{12},
  3475--3482\relax
\mciteBstWouldAddEndPuncttrue
\mciteSetBstMidEndSepPunct{\mcitedefaultmidpunct}
{\mcitedefaultendpunct}{\mcitedefaultseppunct}\relax
\EndOfBibitem
\bibitem[Sturluson \emph{et~al.}(2019)Sturluson, Huynh, Kaija, Laird, Yoon,
  Hou, Feng, Wilmer, Colón, Chung, Siderius, and Simon]{arni}
A.~Sturluson, M.~T. Huynh, A.~R. Kaija, C.~Laird, S.~Yoon, F.~Hou, Z.~Feng,
  C.~E. Wilmer, Y.~J. Colón, Y.~G. Chung, D.~W. Siderius and C.~M. Simon,
  \emph{Molecular Simulation}, 2019, \textbf{45}, 1082--1121\relax
\mciteBstWouldAddEndPuncttrue
\mciteSetBstMidEndSepPunct{\mcitedefaultmidpunct}
{\mcitedefaultendpunct}{\mcitedefaultseppunct}\relax
\EndOfBibitem
\bibitem[Moghadam \emph{et~al.}(2020)Moghadam, Li, Liu, Bueno-Perez, Wang,
  Wiggin, Wood, and Fairen-Jimenez]{moghadam}
P.~Z. Moghadam, A.~Li, X.-W. Liu, R.~Bueno-Perez, S.-D. Wang, S.~B. Wiggin,
  P.~A. Wood and D.~Fairen-Jimenez, \emph{Chem. Sci.}, 2020, \textbf{11},
  8373--8387\relax
\mciteBstWouldAddEndPuncttrue
\mciteSetBstMidEndSepPunct{\mcitedefaultmidpunct}
{\mcitedefaultendpunct}{\mcitedefaultseppunct}\relax
\EndOfBibitem
\bibitem[Rampal \emph{et~al.}(2021)Rampal, Ajenifuja, Tao, Balzer, Cummings,
  Evans, Bueno-Perez, Law, Bolton, Petit, Siperstein, Attfield, Jobson,
  Moghadam, and Fairen-Jimenez]{toolbox}
N.~Rampal, A.~Ajenifuja, A.~Tao, C.~Balzer, M.~S. Cummings, A.~Evans,
  R.~Bueno-Perez, D.~J. Law, L.~W. Bolton, C.~Petit, F.~Siperstein, M.~P.
  Attfield, M.~Jobson, P.~Z. Moghadam and D.~Fairen-Jimenez, \emph{Chem. Sci.},
  2021,  --\relax
\mciteBstWouldAddEndPuncttrue
\mciteSetBstMidEndSepPunct{\mcitedefaultmidpunct}
{\mcitedefaultendpunct}{\mcitedefaultseppunct}\relax
\EndOfBibitem
\bibitem[Getman \emph{et~al.}(2012)Getman, Bae, Wilmer, and Snurr]{ms_mofs}
R.~B. Getman, Y.-S. Bae, C.~E. Wilmer and R.~Q. Snurr, \emph{Chemical Reviews},
  2012, \textbf{112}, 703--723\relax
\mciteBstWouldAddEndPuncttrue
\mciteSetBstMidEndSepPunct{\mcitedefaultmidpunct}
{\mcitedefaultendpunct}{\mcitedefaultseppunct}\relax
\EndOfBibitem
\bibitem[Yang and Zhong(2006)]{ms_Co2_methane}
Q.~Yang and C.~Zhong, \emph{The Journal of Physical Chemistry B}, 2006,
  \textbf{110}, 17776--17783\relax
\mciteBstWouldAddEndPuncttrue
\mciteSetBstMidEndSepPunct{\mcitedefaultmidpunct}
{\mcitedefaultendpunct}{\mcitedefaultseppunct}\relax
\EndOfBibitem
\bibitem[Chung \emph{et~al.}(2014)Chung, Camp, Haranczyk, Sikora, Bury,
  Krungleviciute, Yildirim, Farha, Sholl, and Snurr]{core}
Y.~G. Chung, J.~Camp, M.~Haranczyk, B.~J. Sikora, W.~Bury, V.~Krungleviciute,
  T.~Yildirim, O.~K. Farha, D.~S. Sholl and R.~Q. Snurr, \emph{Chemistry of
  Materials}, 2014, \textbf{26}, 6185--6192\relax
\mciteBstWouldAddEndPuncttrue
\mciteSetBstMidEndSepPunct{\mcitedefaultmidpunct}
{\mcitedefaultendpunct}{\mcitedefaultseppunct}\relax
\EndOfBibitem
\bibitem[Coudert and Fuchs(2016)]{characterization}
F.-X. Coudert and A.~H. Fuchs, \emph{Coordination Chemistry Reviews}, 2016,
  \textbf{307}, 211--236\relax
\mciteBstWouldAddEndPuncttrue
\mciteSetBstMidEndSepPunct{\mcitedefaultmidpunct}
{\mcitedefaultendpunct}{\mcitedefaultseppunct}\relax
\EndOfBibitem
\bibitem[Li \emph{et~al.}(2017)Li, Vermeulen, Malliakas,
  G{\'o}mez-Gualdr{\'o}n, Howarth, Mehdi, Dohnalkova, Browning,
  O{\textquoteright}Keeffe, and Farha]{diego_bottom_up}
P.~Li, N.~A. Vermeulen, C.~D. Malliakas, D.~A. G{\'o}mez-Gualdr{\'o}n, A.~J.
  Howarth, B.~L. Mehdi, A.~Dohnalkova, N.~D. Browning,
  M.~O{\textquoteright}Keeffe and O.~K. Farha, \emph{Science}, 2017,
  \textbf{356}, 624--627\relax
\mciteBstWouldAddEndPuncttrue
\mciteSetBstMidEndSepPunct{\mcitedefaultmidpunct}
{\mcitedefaultendpunct}{\mcitedefaultseppunct}\relax
\EndOfBibitem
\bibitem[Wilmer \emph{et~al.}(2012)Wilmer, Leaf, Lee, Farha, Hauser, Hupp, and
  Snurr]{wilmer}
C.~Wilmer, M.~Leaf, C.~Lee, O.~Farha, B.~Hauser, J.~Hupp and R.~Snurr,
  \emph{Nature Chemistry}, 2012, \textbf{4}, 83--89\relax
\mciteBstWouldAddEndPuncttrue
\mciteSetBstMidEndSepPunct{\mcitedefaultmidpunct}
{\mcitedefaultendpunct}{\mcitedefaultseppunct}\relax
\EndOfBibitem
\bibitem[Bao \emph{et~al.}(2015)Bao, Martin, Simon, Haranczyk, Smit, and
  Deem]{mof_hts}
Y.~Bao, R.~Martin, C.~Simon, M.~Haranczyk, B.~Smit and M.~Deem, \emph{The
  Journal of Physical Chemistry C}, 2015, \textbf{119}, 186--195\relax
\mciteBstWouldAddEndPuncttrue
\mciteSetBstMidEndSepPunct{\mcitedefaultmidpunct}
{\mcitedefaultendpunct}{\mcitedefaultseppunct}\relax
\EndOfBibitem
\bibitem[Li \emph{et~al.}(2016)Li, Chung, and Snurr]{song_hts}
S.~Li, Y.~G. Chung and R.~Q. Snurr, \emph{Langmuir}, 2016, \textbf{32},
  10368--10376\relax
\mciteBstWouldAddEndPuncttrue
\mciteSetBstMidEndSepPunct{\mcitedefaultmidpunct}
{\mcitedefaultendpunct}{\mcitedefaultseppunct}\relax
\EndOfBibitem
\bibitem[Wollmann \emph{et~al.}(2011)Wollmann, Leistner, Stoeck, Grünker,
  Gedrich, Klein, Throl, Grählert, Senkovska, Dreisbach, and Kaskel]{wollman}
P.~Wollmann, M.~Leistner, U.~Stoeck, R.~Grünker, K.~Gedrich, N.~Klein,
  O.~Throl, W.~Grählert, I.~Senkovska, F.~Dreisbach and S.~Kaskel, \emph{Chem.
  Commun.}, 2011, \textbf{47}, 5151--5153\relax
\mciteBstWouldAddEndPuncttrue
\mciteSetBstMidEndSepPunct{\mcitedefaultmidpunct}
{\mcitedefaultendpunct}{\mcitedefaultseppunct}\relax
\EndOfBibitem
\bibitem[Mukherjee and Colón(2021)]{Krish}
K.~Mukherjee and Y.~J. Colón, \emph{Molecular Simulation}, 2021, \textbf{47},
  857--877\relax
\mciteBstWouldAddEndPuncttrue
\mciteSetBstMidEndSepPunct{\mcitedefaultmidpunct}
{\mcitedefaultendpunct}{\mcitedefaultseppunct}\relax
\EndOfBibitem
\bibitem[Shi \emph{et~al.}(2020)Shi, Yang, Deng, Cai, Yan, Liang, Liu, and
  Qiao]{ml_shi}
Z.~Shi, W.~Yang, X.~Deng, C.~Cai, Y.~Yan, H.~Liang, Z.~Liu and Z.~Qiao,
  \emph{Mol. Syst. Des. Eng.}, 2020, \textbf{5}, 725--742\relax
\mciteBstWouldAddEndPuncttrue
\mciteSetBstMidEndSepPunct{\mcitedefaultmidpunct}
{\mcitedefaultendpunct}{\mcitedefaultseppunct}\relax
\EndOfBibitem
\bibitem[Fernandez \emph{et~al.}(2014)Fernandez, Boyd, Daff, Aghaji, and
  Woo]{co2_fer}
M.~Fernandez, P.~G. Boyd, T.~D. Daff, M.~Z. Aghaji and T.~K. Woo, \emph{The
  Journal of Physical Chemistry Letters}, 2014, \textbf{5}, 3056--3060\relax
\mciteBstWouldAddEndPuncttrue
\mciteSetBstMidEndSepPunct{\mcitedefaultmidpunct}
{\mcitedefaultendpunct}{\mcitedefaultseppunct}\relax
\EndOfBibitem
\bibitem[Aghaji \emph{et~al.}(2016)Aghaji, Fernandez, Boyd, Daff, and
  Woo]{agaji}
M.~Z. Aghaji, M.~Fernandez, P.~G. Boyd, T.~D. Daff and T.~K. Woo,
  \emph{European Journal of Inorganic Chemistry}, 2016, \textbf{2016},
  4505--4511\relax
\mciteBstWouldAddEndPuncttrue
\mciteSetBstMidEndSepPunct{\mcitedefaultmidpunct}
{\mcitedefaultendpunct}{\mcitedefaultseppunct}\relax
\EndOfBibitem
\bibitem[Chung \emph{et~al.}(2016)Chung, G{\'o}mez-Gualdr{\'o}n, Li, Leperi,
  Deria, Zhang, Vermeulen, Stoddart, You, Hupp, Farha, and Snurr]{Chung_ml_co2}
Y.~G. Chung, D.~A. G{\'o}mez-Gualdr{\'o}n, P.~Li, K.~T. Leperi, P.~Deria,
  H.~Zhang, N.~A. Vermeulen, J.~F. Stoddart, F.~You, J.~T. Hupp, O.~K. Farha
  and R.~Q. Snurr, \emph{Science Advances}, 2016, \textbf{2}, e1600909\relax
\mciteBstWouldAddEndPuncttrue
\mciteSetBstMidEndSepPunct{\mcitedefaultmidpunct}
{\mcitedefaultendpunct}{\mcitedefaultseppunct}\relax
\EndOfBibitem
\bibitem[Thornton \emph{et~al.}(2017)Thornton, Simon, Kim, Kwon, Deeg, Konstas,
  Pas, Hill, Winkler, Haranczyk, and Smit]{h2_ml}
A.~W. Thornton, C.~M. Simon, J.~Kim, O.~Kwon, K.~S. Deeg, K.~Konstas, S.~J.
  Pas, M.~R. Hill, D.~A. Winkler, M.~Haranczyk and B.~Smit, \emph{Chemistry of
  Materials}, 2017, \textbf{29}, 2844--2854\relax
\mciteBstWouldAddEndPuncttrue
\mciteSetBstMidEndSepPunct{\mcitedefaultmidpunct}
{\mcitedefaultendpunct}{\mcitedefaultseppunct}\relax
\EndOfBibitem
\bibitem[Bobbitt and Snurr(2019)]{bobbitt}
N.~S. Bobbitt and R.~Q. Snurr, \emph{Molecular Simulation}, 2019, \textbf{45},
  1069--1081\relax
\mciteBstWouldAddEndPuncttrue
\mciteSetBstMidEndSepPunct{\mcitedefaultmidpunct}
{\mcitedefaultendpunct}{\mcitedefaultseppunct}\relax
\EndOfBibitem
\bibitem[Pardakhti \emph{et~al.}(2017)Pardakhti, Moharreri, Wanik, Suib, and
  Srivastava]{methane_ml}
M.~Pardakhti, E.~Moharreri, D.~Wanik, S.~L. Suib and R.~Srivastava, \emph{ACS
  Combinatorial Science}, 2017, \textbf{19}, 640--645\relax
\mciteBstWouldAddEndPuncttrue
\mciteSetBstMidEndSepPunct{\mcitedefaultmidpunct}
{\mcitedefaultendpunct}{\mcitedefaultseppunct}\relax
\EndOfBibitem
\bibitem[Fanourgakis \emph{et~al.}(2019)Fanourgakis, Gkagkas, Tylianakis,
  Klontzas, and Froudakis]{ML_methane_fanour}
G.~S. Fanourgakis, K.~Gkagkas, E.~Tylianakis, E.~Klontzas and G.~Froudakis,
  \emph{The Journal of Physical Chemistry A}, 2019, \textbf{123},
  6080--6087\relax
\mciteBstWouldAddEndPuncttrue
\mciteSetBstMidEndSepPunct{\mcitedefaultmidpunct}
{\mcitedefaultendpunct}{\mcitedefaultseppunct}\relax
\EndOfBibitem
\bibitem[Simon \emph{et~al.}(2015)Simon, Mercado, Schnell, Smit, and
  Haranczyk]{xekr}
C.~M. Simon, R.~Mercado, S.~K. Schnell, B.~Smit and M.~Haranczyk,
  \emph{Chemistry of Materials}, 2015, \textbf{27}, 4459--4475\relax
\mciteBstWouldAddEndPuncttrue
\mciteSetBstMidEndSepPunct{\mcitedefaultmidpunct}
{\mcitedefaultendpunct}{\mcitedefaultseppunct}\relax
\EndOfBibitem
\bibitem[Fernandez and Barnard(2016)]{geometrical}
M.~Fernandez and A.~S. Barnard, \emph{ACS Combinatorial Science}, 2016,
  \textbf{18}, 243--252\relax
\mciteBstWouldAddEndPuncttrue
\mciteSetBstMidEndSepPunct{\mcitedefaultmidpunct}
{\mcitedefaultendpunct}{\mcitedefaultseppunct}\relax
\EndOfBibitem
\bibitem[Fernandez \emph{et~al.}(2014)Fernandez, Boyd, Daff, Aghaji, and
  Woo]{rapid_ML_co2}
M.~Fernandez, P.~G. Boyd, T.~D. Daff, M.~Z. Aghaji and T.~K. Woo, \emph{The
  Journal of Physical Chemistry Letters}, 2014, \textbf{5}, 3056--3060\relax
\mciteBstWouldAddEndPuncttrue
\mciteSetBstMidEndSepPunct{\mcitedefaultmidpunct}
{\mcitedefaultendpunct}{\mcitedefaultseppunct}\relax
\EndOfBibitem
\bibitem[Bucior \emph{et~al.}(2019)Bucior, Bobbitt, Islamoglu, Goswami,
  Gopalan, Yildirim, Farha, Bagheri, and Snurr]{energy-based-descriptor}
B.~J. Bucior, N.~S. Bobbitt, T.~Islamoglu, S.~Goswami, A.~Gopalan, T.~Yildirim,
  O.~K. Farha, N.~Bagheri and R.~Q. Snurr, \emph{Mol. Syst. Des. Eng.}, 2019,
  \textbf{4}, 162--174\relax
\mciteBstWouldAddEndPuncttrue
\mciteSetBstMidEndSepPunct{\mcitedefaultmidpunct}
{\mcitedefaultendpunct}{\mcitedefaultseppunct}\relax
\EndOfBibitem
\bibitem[Sturluson \emph{et~al.}(2018)Sturluson, Huynh, York, and
  Simon]{eigen_cages}
A.~Sturluson, M.~T. Huynh, A.~H.~P. York and C.~M. Simon, \emph{ACS Central
  Science}, 2018, \textbf{4}, 1663--1676\relax
\mciteBstWouldAddEndPuncttrue
\mciteSetBstMidEndSepPunct{\mcitedefaultmidpunct}
{\mcitedefaultendpunct}{\mcitedefaultseppunct}\relax
\EndOfBibitem
\bibitem[Befort \emph{et~al.}(2021)Befort, DeFever, Tow, Dowling, and
  Maginn]{befort2021machine}
B.~J. Befort, R.~S. DeFever, G.~M. Tow, A.~W. Dowling and E.~J. Maginn,
  \emph{Machine Learning Directed Optimization of Classical Molecular Modeling
  Force Fields}, 2021\relax
\mciteBstWouldAddEndPuncttrue
\mciteSetBstMidEndSepPunct{\mcitedefaultmidpunct}
{\mcitedefaultendpunct}{\mcitedefaultseppunct}\relax
\EndOfBibitem
\bibitem[Ma \emph{et~al.}(2020)Ma, Colón, and Luo]{TL_colon}
R.~Ma, Y.~J. Colón and T.~Luo, \emph{ACS Applied Materials \& Interfaces},
  2020, \textbf{12}, 34041--34048\relax
\mciteBstWouldAddEndPuncttrue
\mciteSetBstMidEndSepPunct{\mcitedefaultmidpunct}
{\mcitedefaultendpunct}{\mcitedefaultseppunct}\relax
\EndOfBibitem
\bibitem[Anderson \emph{et~al.}(2020)Anderson, Biong, and
  Gómez-Gualdrón]{anderson}
R.~Anderson, A.~Biong and D.~A. Gómez-Gualdrón, \emph{Journal of Chemical
  Theory and Computation}, 2020, \textbf{16}, 1271--1283\relax
\mciteBstWouldAddEndPuncttrue
\mciteSetBstMidEndSepPunct{\mcitedefaultmidpunct}
{\mcitedefaultendpunct}{\mcitedefaultseppunct}\relax
\EndOfBibitem
\bibitem[Sturluson \emph{et~al.}(2021)Sturluson, Raza, McConachie, Siderius,
  Fern, and Simon]{arni_cofs}
A.~Sturluson, A.~Raza, G.~D. McConachie, D.~Siderius, X.~Fern and C.~Simon,
  \emph{ChemRxiv}, 2021,  1--25\relax
\mciteBstWouldAddEndPuncttrue
\mciteSetBstMidEndSepPunct{\mcitedefaultmidpunct}
{\mcitedefaultendpunct}{\mcitedefaultseppunct}\relax
\EndOfBibitem
\bibitem[Uteva \emph{et~al.}(2018)Uteva, Graham, Wilkinson, and
  Wheatley]{active_learning}
E.~Uteva, R.~S. Graham, R.~D. Wilkinson and R.~J. Wheatley, \emph{The Journal
  of Chemical Physics}, 2018, \textbf{149}, 174114\relax
\mciteBstWouldAddEndPuncttrue
\mciteSetBstMidEndSepPunct{\mcitedefaultmidpunct}
{\mcitedefaultendpunct}{\mcitedefaultseppunct}\relax
\EndOfBibitem
\bibitem[Vandermause \emph{et~al.}(2019)Vandermause, Torrisi, Batzner, Xie,
  Sun, Kolpak, and Kozinsky]{vandermause}
J.~Vandermause, S.~B. Torrisi, S.~Batzner, Y.~Xie, L.~Sun, A.~M. Kolpak and
  B.~Kozinsky, \emph{On-the-Fly Active Learning of Interpretable Bayesian Force
  Fields for Atomistic Rare Events}, 2019\relax
\mciteBstWouldAddEndPuncttrue
\mciteSetBstMidEndSepPunct{\mcitedefaultmidpunct}
{\mcitedefaultendpunct}{\mcitedefaultseppunct}\relax
\EndOfBibitem
\bibitem[Santos \emph{et~al.}(2020)Santos, Mehana, Wu, Prodanović, Kang,
  Lubbers, Viswanathan, and Pyrcz]{santos_al}
J.~E. Santos, M.~Mehana, H.~Wu, M.~Prodanović, Q.~Kang, N.~Lubbers,
  H.~Viswanathan and M.~J. Pyrcz, \emph{The Journal of Physical Chemistry C},
  2020, \textbf{124}, 22200--22211\relax
\mciteBstWouldAddEndPuncttrue
\mciteSetBstMidEndSepPunct{\mcitedefaultmidpunct}
{\mcitedefaultendpunct}{\mcitedefaultseppunct}\relax
\EndOfBibitem
\bibitem[Rasmussen and Williams(2006)]{gaussianprocesses}
C.~E. Rasmussen and C.~K.~I. Williams, \emph{Gaussian Processes for Machine
  Learning}, MIT Press, 2006\relax
\mciteBstWouldAddEndPuncttrue
\mciteSetBstMidEndSepPunct{\mcitedefaultmidpunct}
{\mcitedefaultendpunct}{\mcitedefaultseppunct}\relax
\EndOfBibitem
\bibitem[Gopalan \emph{et~al.}(2019)Gopalan, Bucior, Bobbitt, and
  Snurr]{gopalan}
A.~Gopalan, B.~Bucior, N.~Bobbitt and R.~Snurr, \emph{Molecular Physics}, 2019,
  \textbf{117}, 3683--3694\relax
\mciteBstWouldAddEndPuncttrue
\mciteSetBstMidEndSepPunct{\mcitedefaultmidpunct}
{\mcitedefaultendpunct}{\mcitedefaultseppunct}\relax
\EndOfBibitem
\bibitem[Liu and Nocedal(1989)]{lbgfs}
D.~Liu and J.~Nocedal, \emph{Mathematical Programming}, 1989, \textbf{45},
  503--528\relax
\mciteBstWouldAddEndPuncttrue
\mciteSetBstMidEndSepPunct{\mcitedefaultmidpunct}
{\mcitedefaultendpunct}{\mcitedefaultseppunct}\relax
\EndOfBibitem
\bibitem[Pedregosa \emph{et~al.}(2011)Pedregosa, Varoquaux, Gramfort, Michel,
  Thirion, Grisel, Blondel, Prettenhofer, Weiss, Dubourg, Vanderplas, Passos,
  Cournapeau, Brucher, Perrot, and Duchesnay]{scikit-learn}
F.~Pedregosa, G.~Varoquaux, A.~Gramfort, V.~Michel, B.~Thirion, O.~Grisel,
  M.~Blondel, P.~Prettenhofer, R.~Weiss, V.~Dubourg, J.~Vanderplas, A.~Passos,
  D.~Cournapeau, M.~Brucher, M.~Perrot and E.~Duchesnay, \emph{Journal of
  Machine Learning Research}, 2011, \textbf{12}, 2825--2830\relax
\mciteBstWouldAddEndPuncttrue
\mciteSetBstMidEndSepPunct{\mcitedefaultmidpunct}
{\mcitedefaultendpunct}{\mcitedefaultseppunct}\relax
\EndOfBibitem
\bibitem[MacKay(1992)]{al_macay}
D.~J.~C. MacKay, \emph{Neural Computation}, 1992, \textbf{4}, 590--604\relax
\mciteBstWouldAddEndPuncttrue
\mciteSetBstMidEndSepPunct{\mcitedefaultmidpunct}
{\mcitedefaultendpunct}{\mcitedefaultseppunct}\relax
\EndOfBibitem
\bibitem[Seo \emph{et~al.}(2000)Seo, Wallat, Graepel, and Obermayer]{seo_gp}
S.~Seo, M.~Wallat, T.~Graepel and K.~Obermayer, Proceedings of the
  International Joint Conference on Neural Networks, 2000, pp. 241 -- 246
  vol.3\relax
\mciteBstWouldAddEndPuncttrue
\mciteSetBstMidEndSepPunct{\mcitedefaultmidpunct}
{\mcitedefaultendpunct}{\mcitedefaultseppunct}\relax
\EndOfBibitem
\bibitem[Maurin \emph{et~al.}(2005)Maurin, Llewellyn, and Bell]{gcmc}
G.~Maurin, P.~L. Llewellyn and R.~G. Bell, \emph{The Journal of Physical
  Chemistry B}, 2005, \textbf{109}, 16084--16091\relax
\mciteBstWouldAddEndPuncttrue
\mciteSetBstMidEndSepPunct{\mcitedefaultmidpunct}
{\mcitedefaultendpunct}{\mcitedefaultseppunct}\relax
\EndOfBibitem
\bibitem[Snurr \emph{et~al.}(1993)Snurr, Bell, and Theodorou]{gcmc2}
R.~Q. Snurr, A.~T. Bell and D.~N. Theodorou, \emph{The Journal of Physical
  Chemistry}, 1993, \textbf{97}, 13742--13752\relax
\mciteBstWouldAddEndPuncttrue
\mciteSetBstMidEndSepPunct{\mcitedefaultmidpunct}
{\mcitedefaultendpunct}{\mcitedefaultseppunct}\relax
\EndOfBibitem
\bibitem[Dubbeldam \emph{et~al.}(2016)Dubbeldam, Calero, Ellis, and
  Snurr]{raspa}
D.~Dubbeldam, S.~Calero, D.~E. Ellis and R.~Q. Snurr, \emph{Molecular
  Simulation}, 2016, \textbf{42}, 81--101\relax
\mciteBstWouldAddEndPuncttrue
\mciteSetBstMidEndSepPunct{\mcitedefaultmidpunct}
{\mcitedefaultendpunct}{\mcitedefaultseppunct}\relax
\EndOfBibitem
\bibitem[Eggimann \emph{et~al.}(2014)Eggimann, Sunnarborg, Stern, Bliss, and
  Siepmann]{trappe}
B.~Eggimann, A.~Sunnarborg, H.~Stern, A.~Bliss and J.~Siepmann, \emph{Molecular
  Simulation}, 2014, \textbf{40}, 101--105\relax
\mciteBstWouldAddEndPuncttrue
\mciteSetBstMidEndSepPunct{\mcitedefaultmidpunct}
{\mcitedefaultendpunct}{\mcitedefaultseppunct}\relax
\EndOfBibitem
\bibitem[Lennard-Jones(1931)]{Lennard_Jones_1931}
J.~E. Lennard-Jones, \emph{Proceedings of the Physical Society}, 1931,
  \textbf{43}, 461--482\relax
\mciteBstWouldAddEndPuncttrue
\mciteSetBstMidEndSepPunct{\mcitedefaultmidpunct}
{\mcitedefaultendpunct}{\mcitedefaultseppunct}\relax
\EndOfBibitem
\bibitem[Rappe \emph{et~al.}(1992)Rappe, Casewit, Colwell, Goddard, and
  Skiff]{UFF}
A.~K. Rappe, C.~J. Casewit, K.~S. Colwell, W.~A. Goddard and W.~M. Skiff,
  \emph{Journal of the American Chemical Society}, 1992, \textbf{114},
  10024--10035\relax
\mciteBstWouldAddEndPuncttrue
\mciteSetBstMidEndSepPunct{\mcitedefaultmidpunct}
{\mcitedefaultendpunct}{\mcitedefaultseppunct}\relax
\EndOfBibitem
\bibitem[{Lorentz}(1881)]{lorentz}
H.~A. {Lorentz}, \emph{Annalen der Physik}, 1881, \textbf{248}, 127--136\relax
\mciteBstWouldAddEndPuncttrue
\mciteSetBstMidEndSepPunct{\mcitedefaultmidpunct}
{\mcitedefaultendpunct}{\mcitedefaultseppunct}\relax
\EndOfBibitem
\bibitem[McKay \emph{et~al.}(1979)McKay, Beckman, and Conover]{LHS}
M.~D. McKay, R.~J. Beckman and W.~J. Conover, \emph{Technometrics}, 1979,
  \textbf{21}, 239--245\relax
\mciteBstWouldAddEndPuncttrue
\mciteSetBstMidEndSepPunct{\mcitedefaultmidpunct}
{\mcitedefaultendpunct}{\mcitedefaultseppunct}\relax
\EndOfBibitem
\bibitem[Walton and Sholl(2015)]{IAST_50}
K.~S. Walton and D.~S. Sholl, \emph{AIChE Journal}, 2015, \textbf{61},
  2757--2762\relax
\mciteBstWouldAddEndPuncttrue
\mciteSetBstMidEndSepPunct{\mcitedefaultmidpunct}
{\mcitedefaultendpunct}{\mcitedefaultseppunct}\relax
\EndOfBibitem
\bibitem[Simpson \emph{et~al.}(2001)Simpson, Peplinski, Koch, and
  Allen]{novel_priors}
T.~Simpson, J.~Peplinski, P.~Koch and J.~Allen, \emph{Engineering with
  Computers}, 2001, \textbf{17}, 129--150\relax
\mciteBstWouldAddEndPuncttrue
\mciteSetBstMidEndSepPunct{\mcitedefaultmidpunct}
{\mcitedefaultendpunct}{\mcitedefaultseppunct}\relax
\EndOfBibitem
\end{mcitethebibliography}
\bibliographystyle{rsc} 

\twocolumn[
  \begin{@twocolumnfalse}
  \noindent\LARGE{\textbf{Supporting Information for ---
Sequential Design of Adsorption Simulations in Metal-Organic Frameworks
}} \\
 \vspace{0.01cm} \\
 \noindent\large{Krishnendu Mukherjee, Alexander W. Dowling, and Yamil J. Col\'{o}n} \\
 \end{@twocolumnfalse}
\nopagebreak
]
\nopagebreak
\renewcommand*\rmdefault{bch}\normalfont\upshape
\rmfamily
\section*{}
\vspace{-5cm}
\nopagebreak

\begin{figure*}
\centering
  \includegraphics[height=5.0cm]{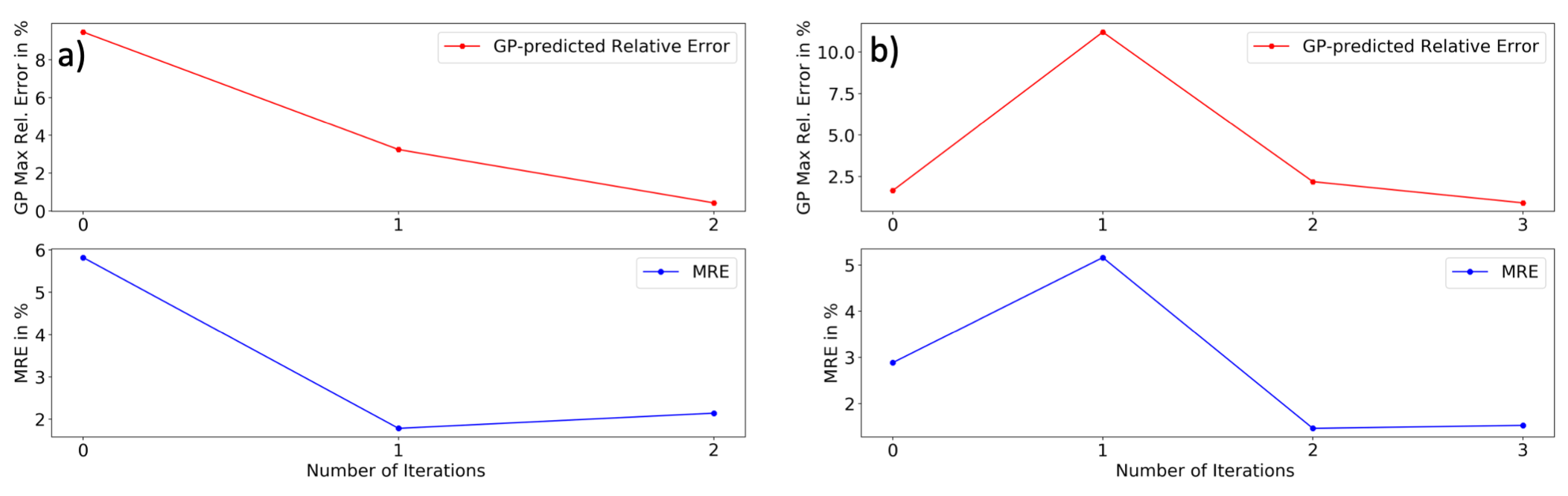}
  \caption{Maximum GP relative error and MRE (in \%) with respect to number of AL iterations for a) methane adsorption, b) CO\textsubscript{2} adsorption for boundary-informed prior. This was done for AL for simulating adsorption isotherm in Cu-BTC at a temperature of 300 K (section 3.1 and 3.2). We find here that methane adsorption took only 2 number of AL iteration to converge at a 2 \% while CO\textsubscript{2} adsorption took 3 iterations of AL}
  \label{fgr:Max_GP_MRE_1f}
\end{figure*}

\begin{figure*}
\centering
  \includegraphics[height=6.0cm]{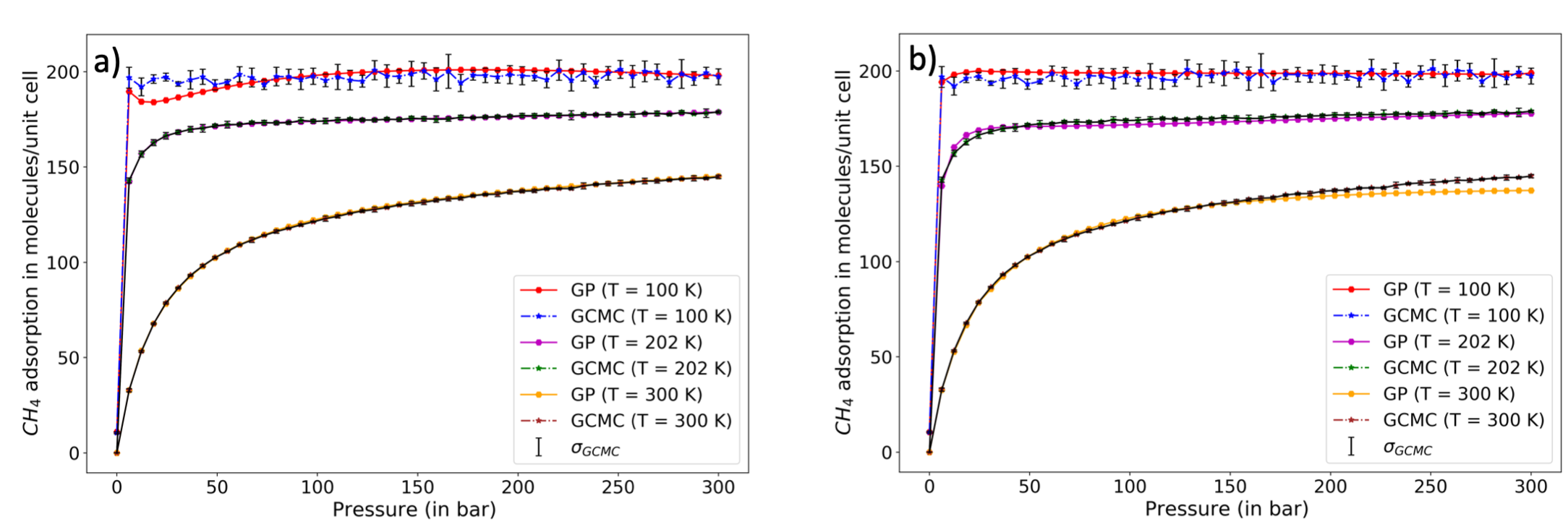}
  \caption{Comparison of GP-predicted methane uptake with GCMC simulation predicted for pressure range of 10\textsuperscript{-6} to 100 bar, at temperature of 100 K, 202 K and 300 K for a) linear-spaced LHS prior, and b) log-spaced LHS prior. We observe here that linear-spaced prior had good agreement with GCMC results at high temperature but for low temperature (at 100 K), there was error at the low-pressure region. For log-spaced prior we observe that the final model has good accuracy at low temperature but for high temperature of 300 K, there is deviation with GCMC simulation at high pressure}
  \label{fgr:CH4_2f_others}
\end{figure*}

\begin{figure*}
\centering
  \includegraphics[height=4.0cm]{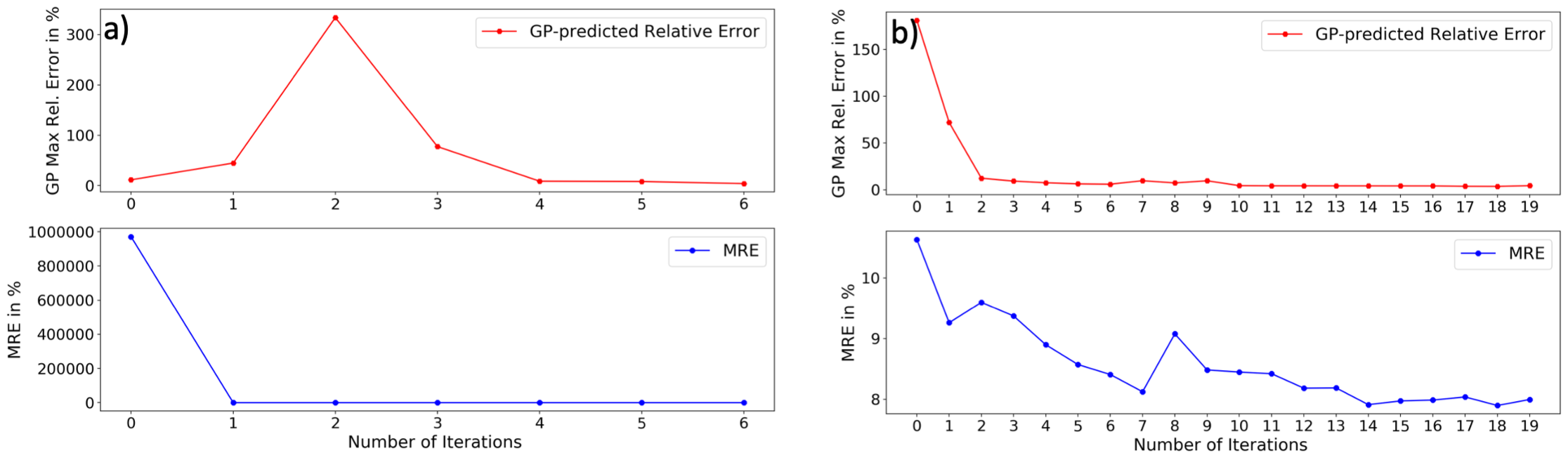}
  \caption{Maximum GP relative error and MRE (in \%) with respect to number of AL iterations for methane adsorption in two features. The pressure range here is 10\textsuperscript{-6} to 300 bar, and temperature range of 100 K to 300 K. The plots are for a) linear-spaced LHS, b) log-spaced LHS. We find a very high maximum GP relative error for linear-spaced prior and a jump to higher GP relative error and a coming back to below 2 \%. This same error for log-spaced prior goes much smoothly to below 2 \% but take a greater number of iterations}
  \label{fgr:max_GP_MRE_methane_2f}
\end{figure*}

\begin{figure*}
\centering
  \includegraphics[height=6.0cm]{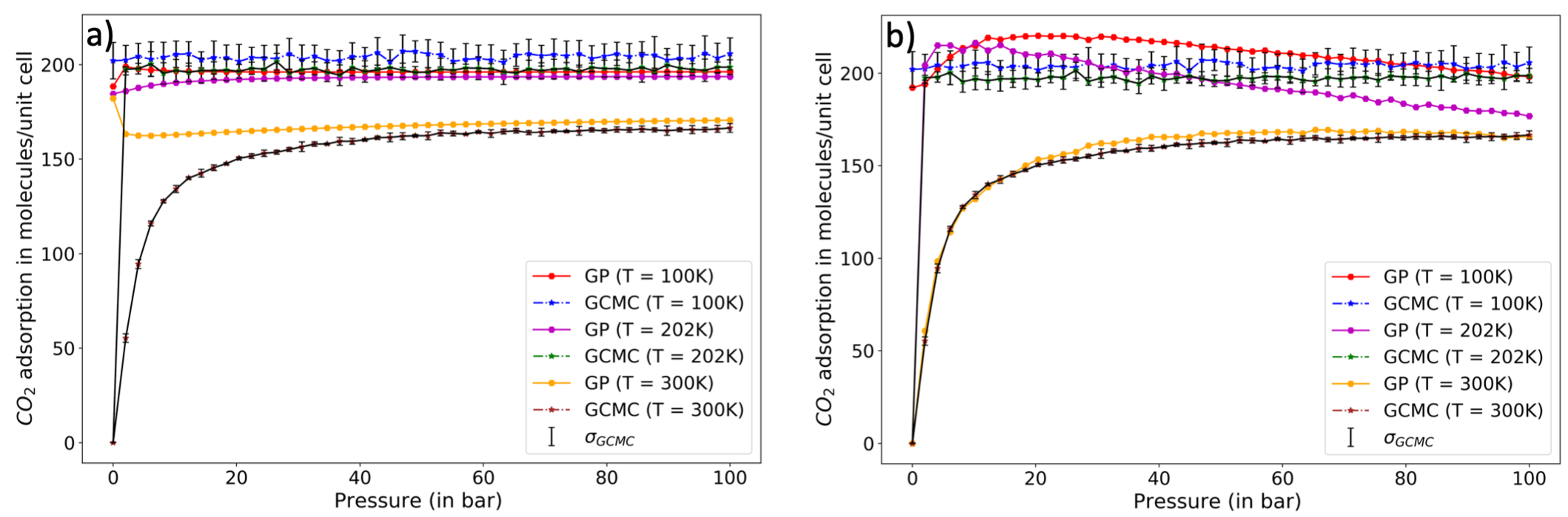}
  \caption{Comparison of GP-predicted CO\textsubscript{2} uptake with GCMC simulation predicted for pressure range of 10\textsuperscript{-6} to 100 bar, at temperature of 100 K, 202 K and 300 K for a) linear-spaced LHS prior, and b) log-spaced LHS prior. We find a high disagreement with GCMC for linear-spaced prior for all the temperature at low-pressure region. For log-spaced there is a high error at low temperature and this error is more pronounced at the high-pressure region}
  \label{fgr:CO2_2f_others}
\end{figure*}

\begin{figure*}
\centering
  \includegraphics[height=5.0cm]{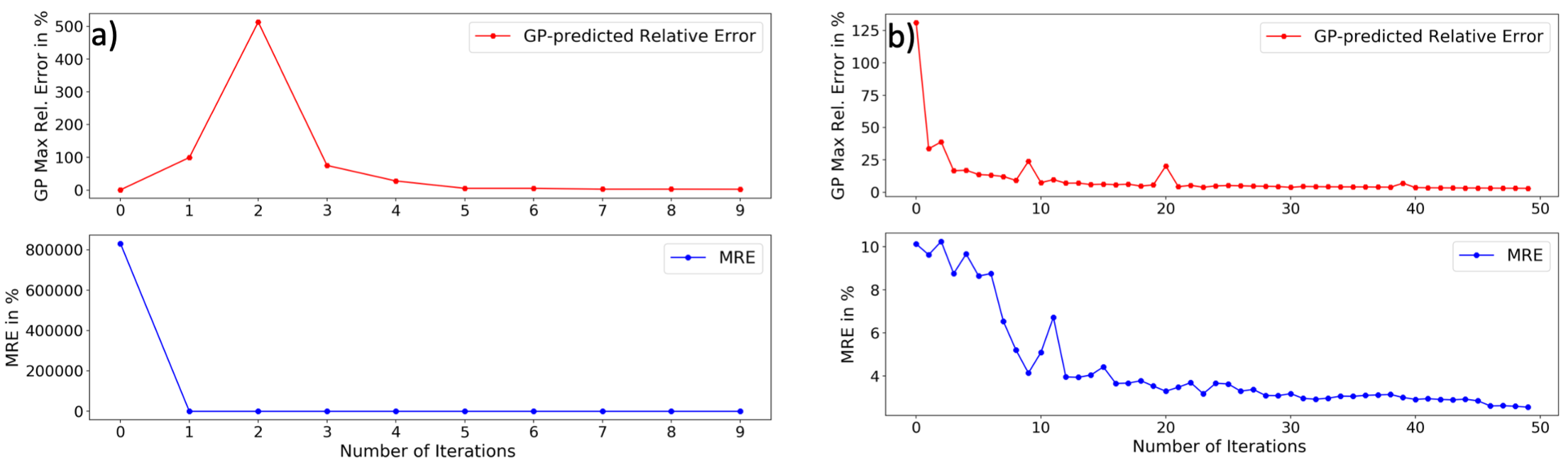}
  \caption{Maximum GP relative error and MRE (in \%) with respect to number of AL iterations for 2 features for CO\textsubscript{2} adsorption. The pressure range here is 10\textsuperscript{-6} to 300 bar, and temperature range is 100 K to 300 K. The plots are for a) linear-spaced LHS, b) log-spaced LHS. Please note the convergence limit for CO\textsubscript{2} adsorption for two features was set to 3 \% while for methane it was 2 \%. Here also we observe a spike in GP maximum relative error for linear-spaced prior and then decreasing to go below 3 \% limit. For log-spaced prior we a smooth decline except few points of rise. The log-spaced prior takes considerably a greater number of iterations to converge but has a comparable final MRE of 2.64 \% while linear one has 2.79 \%}
  \label{fgr:Max_GP_MRE_CO2_2f}
\end{figure*}

\begin{figure*}
\centering
  \includegraphics[height=12.0cm]{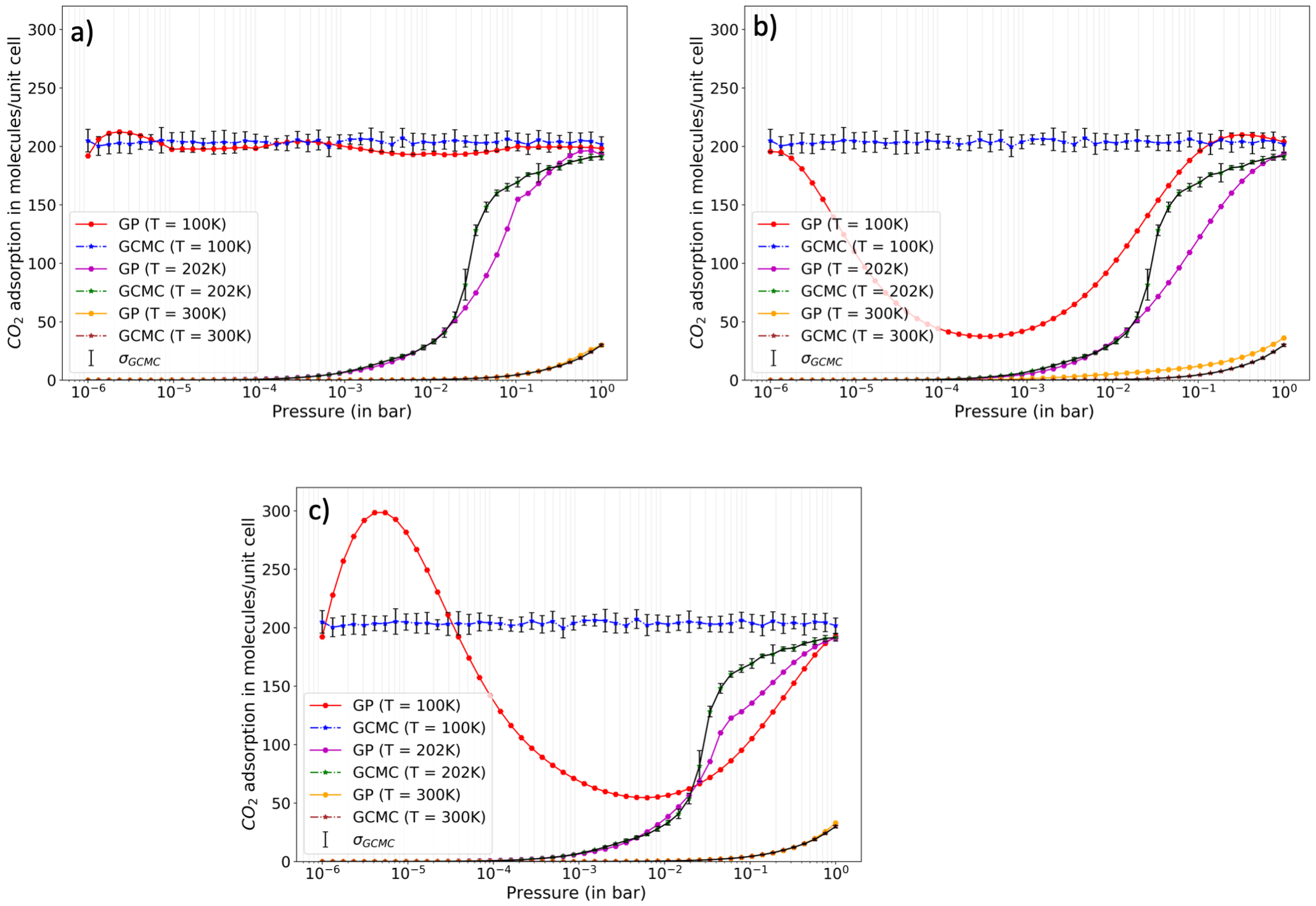}
  \caption{CO\textsubscript{2} uptake comparison between GP and GCMC simulation in Cu-BTC at low pressure range of 10–6 to 1 bar for different priors, a) boundary-informed prior, b) linear-spaced LHS prior, and c) log-spaced LHS prior. We find here that boundary performs quite well for at temperature except at the isotherm rise region. Both log-spaced and linear-spaced prior have high error at lowest temperature. Also, at a very low pressure (10\textsuperscript{–6} to 10\textsuperscript{–3} bar) the errors are very high. Since adsorption is basically nil in this zone, the GP has a very uncertainty in prediction. Log-spaced prior performs better than linear-spaced prior at higher temperature. Linear spaced has a poor performance for all the three temperature points}
  \label{fgr:CO2_lowP_uptake}
\end{figure*}

\begin{figure*}
\centering
  \includegraphics[height=12.0cm]{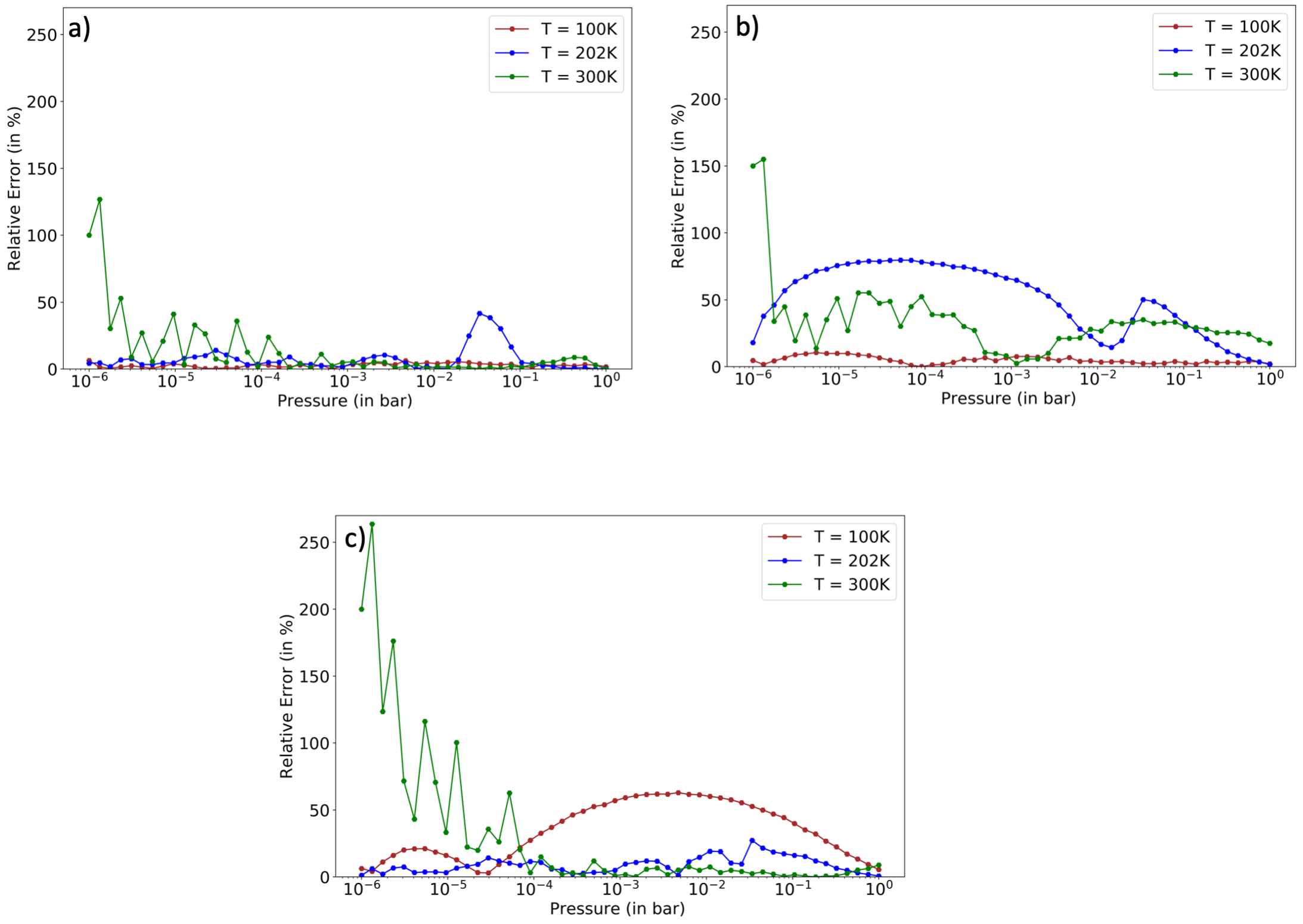}
  \caption{Relative error (in \%) comparison between GP and GCMC simulation in Cu-BTC at low pressure range of 10\textsuperscript{–6} to 1 bar for different priors for CO\textsubscript{2} adsorption, a) boundary-informed prior, b) linear-spaced LHS prior, and c) log-spaced LHS prior. We find boundary-informed prior had a lower relative error throughout all the temperature point. Boundary-informed prior had a few points close to and over 50 \% relative error at the highest temperature of 300 K. The error decreases for lower temperature. This error might be because adsorption at this pressure range and at high temperature is almost zero and there is a high fluctuation even at the GCMC simulation (refer table 5). Linear-spaced prior had considerably higher relative error, but it does decreases with temperature. Log-spaced prior had a very high relative error at 300 K in the further low-pressure change but the relative error decreases as we decrease the temperature}
  \label{fgr:CO2_lowP_MRE}
\end{figure*}

\end{document}